\documentclass[aps, superscriptaddress, twocolumn, longbibliography]{revtex4-1}

\usepackage{amsmath,amssymb}
\usepackage{graphicx}
\usepackage{hyperref}
\usepackage{enumitem}

\begin{document}
\title{Latent geometry and dynamics of proximity networks}
\author{Fragkiskos Papadopoulos}
\email{f.papadopoulos@cut.ac.cy}
\affiliation{Department of Electrical Engineering, Computer Engineering and Informatics, Cyprus University of Technology, 33 Saripolou Street, 3036 Limassol, Cyprus}
\author{Marco Antonio Rodr\'iguez Flores}
\affiliation{Department of Electrical Engineering, Computer Engineering and Informatics, Cyprus University of Technology, 33 Saripolou Street, 3036 Limassol, Cyprus}

\date{\today}
\begin{abstract}
Proximity networks are time-varying graphs representing the closeness among humans moving in a physical space. Their properties have been extensively studied in the past decade as they critically affect the behavior of spreading phenomena and the performance of routing algorithms. Yet, the mechanisms responsible for their observed characteristics remain elusive. Here, we show that many of the observed properties of proximity networks emerge \emph{naturally} and \emph{simultaneously} in a simple latent space network model, called \emph{dynamic-$\mathbb{S}^{1}$}. The dynamic-$\mathbb{S}^{1}$ does not model node mobility directly, but captures the connectivity in each snapshot---each snapshot in the model is a realization of the $\mathbb{S}^{1}$ model of traditional complex networks, which is isomorphic to hyperbolic geometric graphs. By forgoing the motion component the model facilitates mathematical analysis, allowing us to prove the contact, inter-contact and weight distributions. We show that these distributions are power laws in the thermodynamic limit with exponents lying within the ranges observed in real systems. Interestingly, we find that \emph{network temperature} plays a central role in network dynamics, dictating the exponents of these distributions, the time-aggregated agent degrees, and the formation of unique and recurrent components. Further, we show that paradigmatic epidemic and rumor spreading processes perform similarly in real and modeled networks. The dynamic-$\mathbb{S}^{1}$ or extensions of it may apply to other types of time-varying networks and constitute the basis of maximum likelihood estimation methods that infer the node coordinates and their evolution in the latent spaces of real systems.
\end{abstract}

\maketitle

%%%%%%%%%%%%%%%%%%%%%%%%%%%%%%%%%%%%%%%%%%%%%%%%%%%%%%%%%%%%%%%%%%%%%%%
\section{Introduction}
\label{sec:intro}

Understanding the time-varying proximity patterns among humans in a physical space is important in various contexts. These include the analysis and containment of spreading phenomena, like respiratory transmitted diseases, the design of routing algorithms for mobile networks, and the understanding of social relationships and influence~\cite{BarratOverview2015, HolmeStructure2016, HolmeVacc2016, hui_paper, chaintreau_paper, thomas_paper, MITData, FFData}. To this end, proximity networks have been captured in different environments~\cite{chaintreau_paper, HospitalData, PrimaryData, HighSchoolData, OfficeData, ConferenceData, MITData, FFData}. Each snapshot in these networks corresponds to an observation interval, which typically spans a few seconds to several minutes depending on the devices used to collect the data. The agents (nodes) in each snapshot are individuals and an edge between two agents means that they are within proximity range. 

At the finest granularity level an edge between two agents represents a close-range face-to-face proximity (up to $1.5$~m, detected using wearable sensors).  Such networks have been captured over the period of few days or weeks in different closed settings, such as hospitals, schools, scientific conferences and workplaces~\cite{HospitalData, PrimaryData, HighSchoolData, ConferenceData, OfficeData}. The main motivation for obtaining these data has emerged in epidemiological studies of infectious diseases. Other proximity networks have been captured for longer periods of time (months) and over larger areas, such as university campuses, using Bluetooth sensing or WiFi tracking~\cite{chaintreau_paper, MITData, FFData}. These methods yield information only on proximity at a range, e.g., up to $10$~m using Bluetooth devices and up to $40$~m or more using WiFi tracking~\cite{MITData, FFData, DartmouthData}. Thus, proximity in these networks does not imply face-to-face interaction. The collection of these data has been motivated by research in mobile networking~\cite{hui_paper, chaintreau_paper, thomas_paper} and social studies~\cite{MITData, FFData}.

Irrespectively of the context, measurement period, and measurement method, different proximity networks have been shown to exhibit similar statistical properties~\cite{BarratOverview2015, StarniniDevices2017, chaintreau_paper, thomas_paper}. The most widely studied properties are the aggregated---obtained by considering the samples from all pairs of nodes together---distributions of contact and inter-contact durations. The former is the distribution of time that a pair of nodes spends in contact, i.e., remains within proximity range, while the latter is the distribution of time separating two contacts between the same pair of nodes. These metrics are important in determining the capacity and delay of a network, and the dynamics of spreading processes~\cite{ContiOppurtunisticOverview, vazquez2007, timo2009, machens2013, gauvin2013}. It has been found that both of these distributions are broad in real data and compatible with power laws, $P(t) \propto t^{-\gamma}$, with or without exponential cutoffs~\cite{hui_paper, chaintreau_paper, thomas_paper, StarniniDevices2017}. Studies have reported exponents $\gamma \ge 2$ for contact  durations~\cite{SPcontactexp, Scherrer2008} and $\gamma \in (1, 2)$ for inter-contact durations~\cite{hui_paper, chaintreau_paper, Partners2011, HighSchoolData2}. Further, it has been shown that aggregated power laws can emerge from pairwise distributions that are either power-laws, exponentials or log-normals, with the latter two better fitting most pairwise inter-contact durations in real data~\cite{pairwise1, pairwise2, pairwise3}.  Another property of interest is the distribution of the total duration of contacts between two agents throughout the observation period, called weight distribution~\cite{StarniniDevices2017, gauvin2013, memory1}. The aggregated weight distribution is also roughly compatible with power laws~\cite{StarniniDevices2017}, while an exponent $\gamma=1.4$ has been reported for this distribution in the contact network of high school students~\cite{HighSchoolData2}.

These and other distinctive features of real proximity networks can be well reproduced by minimal models of mobile interacting agents~\cite{Starnini2013, StarniniDevices2017, flores2018}. Minimal models, i.e., models that reproduce many of the observed properties under minimal assumptions, are crucial for generating realistic synthetic networks and understanding the mechanisms that are responsible for the observed behaviors. In particular, the recently developed Force-Directed Motion (FDM) model~\cite{flores2018} utilizes the idea of a latent metric space where the agents reside, and where the distance $d$ between two agents abstracts their \emph{similarity}. Attractive forces that decrease exponentially with the similarity distance direct the agents' motion towards other agents in the physical space, and determine the duration of their interactions. One can also consider the \emph{effective distance} between two agents, $\chi=d/(\kappa \kappa')$, where $\kappa$ and $\kappa'$ are the agents' expected degrees per snapshot, abstracting their \emph{popularity}~\cite{Papadopoulos2012}. In this case, dissimilar agents can still be attracted by strong forces if their popularities are high. The FDM casts the problem of modeling proximity networks as an $N$-body problem akin to molecular dynamics~\cite{schlick2010molecular}. However, mathematically proving the properties of generated networks by the FDM is not straightforward, and the model has been so far studied only in simulations.

\begin{figure}
\includegraphics[width=3.55in]{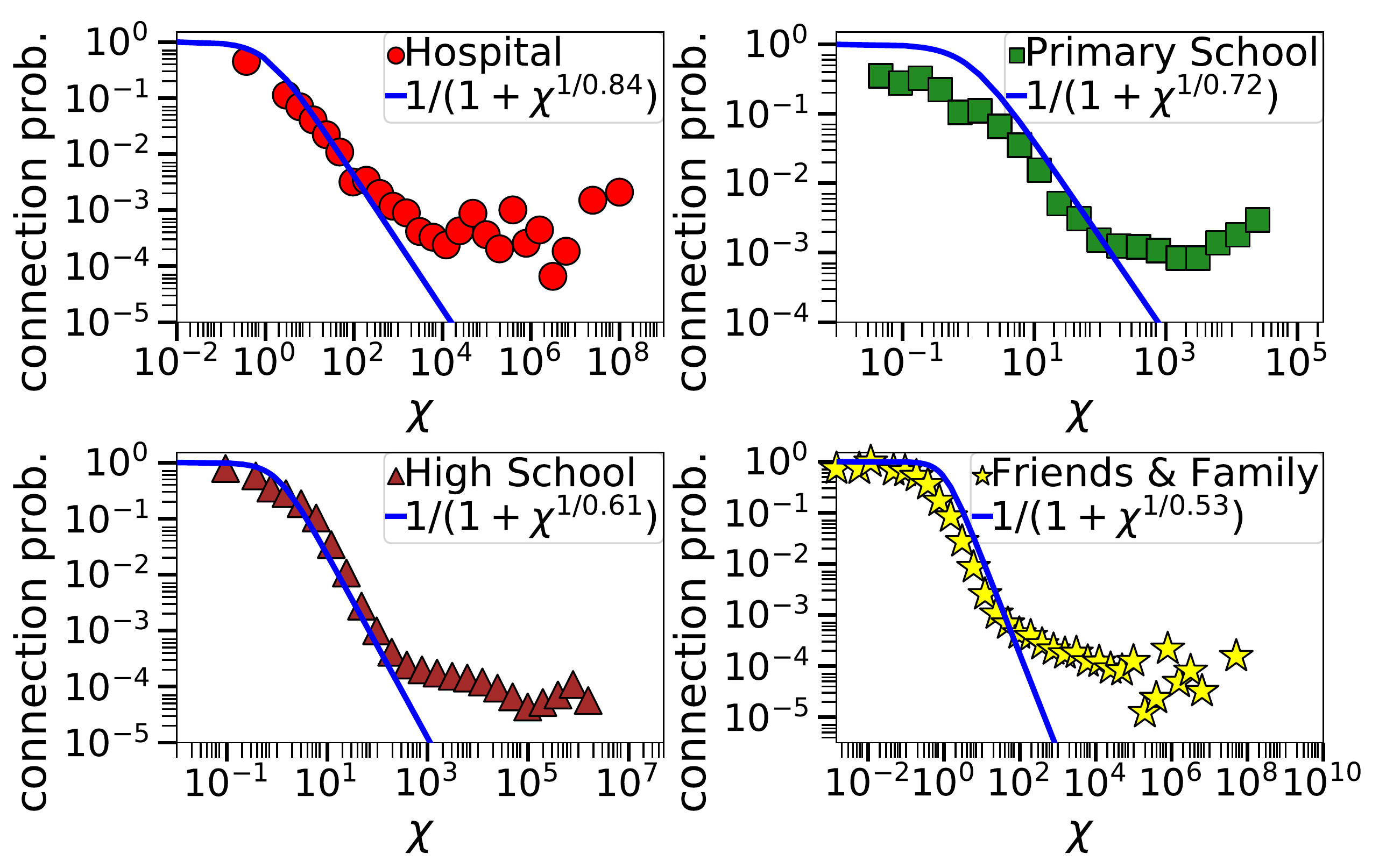}
\caption{Probability that two agents are connected in a snapshot as a function of their effective distance $\chi$ in FDM-simulated counterparts of the hospital, primary school and high school face-to-face interaction networks~\cite{HospitalData, PrimaryData, HighSchoolData}; and of the Friends~\&~Family proximity network~\cite{FFData}. The simulations are performed as in~\cite{flores2018}, while the connection probabilities are computed excluding agents that are inactive~\cite{flores2018} in each snapshot. The solid lines are Fermi-Dirac connection probabilities with temperatures $T=0.84, 0.72, 0.61, 0.53$, corresponding respectively to the temperatures of the hospital, primary school, high school and Friends~\&~Family (Sec.~\ref{sec:modeled_nets}).
\label{fig:fermi}}
\end{figure}

The FDM has been inspired by the $\mathbb{S}^{1}$ model of traditional (non-mobile) complex networks~\cite{Krioukov2010, Serrano2008}. In the $\mathbb{S}^{1}$, nodes are also separated by effective distances $\chi$, and are connected with the \emph{Fermi-Dirac} connection probability $p(\chi)=1/(1+\chi^{1/T})$, where $T \in (0,1)$ is the \emph{network temperature}, controlling clustering~\cite{Dorogovtsev10-book} in the network. The $\mathbb{S}^{1}$ is isomorphic to hyperbolic geometric graphs~\cite{Krioukov2010}. It can generate network snapshots that possess many of the common structural properties of real networks, including heterogeneous or homogeneous degree distributions, strong clustering, and the small-world property~\cite{Serrano2008, Krioukov2010, Papadopoulos2012}. Fig.~\ref{fig:fermi} shows the probability that two agents are connected in a snapshot of FDM-simulated networks as a function of their effective distance. Interestingly, we see that this probability resembles qualitatively the Fermi-Dirac connection probability in the $\mathbb{S}^{1}$ model, even though this form of connection probability is \emph{not} enforced into the FDM. Specifically, we see in Fig.~\ref{fig:fermi} that the connection probability in the FDM has a smooth step-like form, where connection probabilities at small distances are orders of magnitude larger than connection probabilities at large distances.

Motivated by the observation in Fig.~\ref{fig:fermi}, here we consider a simple latent space model for human proximity networks, where each snapshot is a realization of the $\mathbb{S}^{1}$ model. We call this model \emph{dynamic-$\mathbb{S}^{1}$} and show that it \emph{simultaneously} reproduces many of the observed properties of real systems. The dynamic-$\mathbb{S}^{1}$ does not model node mobility directly, but captures the connectivity in each snapshot. By forgoing the motion component it facilitates mathematical analysis, allowing us to prove the contact, inter-contact and weight distributions. We show that these distributions are power laws in the thermodynamic limit, with exponents $2+T$, $2-T$ and $1+T$, respectively, where $T \in (0,1)$ is the temperature in the Fermi-Dirac connection probability. These exponents are within the ranges observed in real systems. We also show that temperature controls the agents' time-aggregated degrees and the formation of unique and recurrent components~\cite{flores2018}. Additionally, we consider paradigmatic epidemic and rumor spreading processes~\cite{sis_ref, Daley1965} and find that they perform remarkably similar in real and modeled networks. 

The rest of the paper is organized as follows. In Sec.~\ref{sec:S1} we review the $\mathbb{S}^{1}$ model. In Sec.~\ref{sec:dynamic_S1} we introduce the dynamic-$\mathbb{S}^{1}$. In Sec.~\ref{sec:real_vs_modeled} we juxtapose the properties of modeled and real networks. In Sec.~\ref{sec:processes} we compare the performance of epidemic and rumor spreading processes running on them. In Sec.~\ref{sec:analysis} we mathematically analyze the main properties of the model. In Sec.~\ref{sec:components_vs_T} we elucidate the crucial role of temperature in the formation of components. Finally, in Sec.~\ref{sec:conclusion} we conclude the paper with future work directions.
 
%%%%%%%%%%%%%%%%%%%%%%%%%%%%%%%%%%%%%%%%%%%%%%%%%%%%%%%%%%%%%%%%%%%%%%%
\section{$\mathbb{S}^{1}$ model}
\label{sec:S1}

In the $\mathbb{S}^{1}$ model~\cite{Krioukov2010} each node has latent (or hidden) variables $\kappa, \theta$. The latent variable $\kappa$ is proportional to the node's expected degree in the resulting network. The latent variable $\theta$ is the angular similarity coordinate of the node on a circle of radius $R=N/2\pi$, where $N$ is the total number of nodes. To construct a network with the model that has size $N$, average node degree $\bar{k}$, and temperature $T \in (0,1)$, we perform the following steps:
\begin{enumerate}
\item[(1)] coordinate assignment: for each node $i=1, 2,\ldots,N$, sample its angular coordinate $\theta_i$ uniformly at random from $[0, 2\pi]$, and its degree variable $\kappa_i$ from a probability density function (PDF) $\rho(\kappa)$;
\item[(2)] creation of edges: connect every pair of nodes $i, j$ with the Fermi-Dirac connection probability
\begin{align}
\label{eq:p_s1}
p(\chi_{ij})=\frac{1}{1+\chi_{ij}^{1/T}}.
\end{align}
\end{enumerate}
In the last expression, $\chi_{ij}$ is the effective distance between nodes $i$ and $j$,
\begin{align}
\label{eq:chi}
\chi_{ij} = \frac{R \Delta\theta_{ij}}{\mu \kappa_i \kappa_j},
\end{align}
where $\Delta \theta_{ij}=\pi - | \pi -|\theta_i - \theta_j||$. Parameter $\mu$ in~(\ref{eq:chi}) is derived from the condition that the expected degree in the network is indeed $\bar{k}$, yielding
\begin{align}
\label{eq:mu}
\mu=\frac{\bar{k}\sin{(T \pi)}}{2\bar{\kappa}^2 T \pi},
\end{align}
where $\bar{\kappa} = \int \kappa \rho(\kappa) \mathrm{d} \kappa$. The expected degree of a node with latent variable $\kappa$ is~\cite{Krioukov2010}
\begin{align}
\label{eq:kappa}
\bar{k}(\kappa)= \frac{\bar{k}}{\bar{\kappa}}\kappa.
\end{align}
For sparse networks ($\bar{k} \ll N$) the resulting degree distribution $P(k)$ has a similar functional form as $\rho(\kappa)$~\cite{Boguna2003}. For instance, a power law degree distribution with exponent $\gamma > 2$ is obtained if $\rho(\kappa) \propto \kappa^{-\gamma}$, while a Poisson degree distribution with mean $\bar{k}$ is obtained if $\rho(\kappa)=\delta(\kappa-\bar{k})$, where $\delta(x)$ is  the Dirac delta function~\cite{Boguna2003, Serrano2008}.  Smaller values of  the temperature $T$ favor connections at smaller effective distances and increase the average clustering~\cite{Dorogovtsev10-book} in the network, which is maximized at $T=0$, and nearly
linearly decreases to zero with $T \in [0, 1)$. At $T \to 0$ the connection probability in~(\ref{eq:p_s1}) becomes the step function $p(\chi_{ij}) \to 1$ if $\chi_{ij} < 1$, and $p(\chi_{ij}) \to 0$ if $\chi_{ij}  > 1$.  

%%%%%%%%%%%%%%%%%%%%%%%%%%%%%%%%%%%%%%%%%%%%%%%%%%%%%%%%%%%%%%%%%%%%%%%
\section{Dynamic-$\mathbb{S}^{1}$}
\label{sec:dynamic_S1}

The dynamic-$\mathbb{S}^{1}$ models a sequence of network snapshots, $G_t$, $t=1,\ldots, \tau$, where $\tau$ is the total number of time slots. Each snapshot is a realization of the $\mathbb{S}^{1}$ model. Therefore, there are $N$ agents that are assigned latent variables $\kappa, \theta$ as in the $\mathbb{S}^{1}$ model, which remain fixed in all time slots. The temperature $T$ is also fixed, while each snapshot $G_t$ is allowed to have a different average degree $\bar{k}_t$. Thus, the model parameters are $N, \tau, \rho(\kappa)$, $T$, and $\bar{k}_t, t=1, \ldots, \tau$.  The snapshots are generated according to the following simple rules:
\begin{enumerate}
\item[(1)] at each time step $t=1, \ldots, \tau$, snapshot $G_t$ starts with $N$ disconnected nodes, while $\bar{k}$ in Eq.~(\ref{eq:mu}) is set equal to $\bar{k}_t$;
\item[(2)] each pair of nodes $i, j$ connects with probability given by Eq.~(\ref{eq:p_s1});
\item[(3)] at time $t+1$, all the edges in snapshot $G_t$ are deleted and the process starts over again to generate snapshot $G_{t+1}$.
\end{enumerate}

\begin{figure*}
\includegraphics[width=17cm]{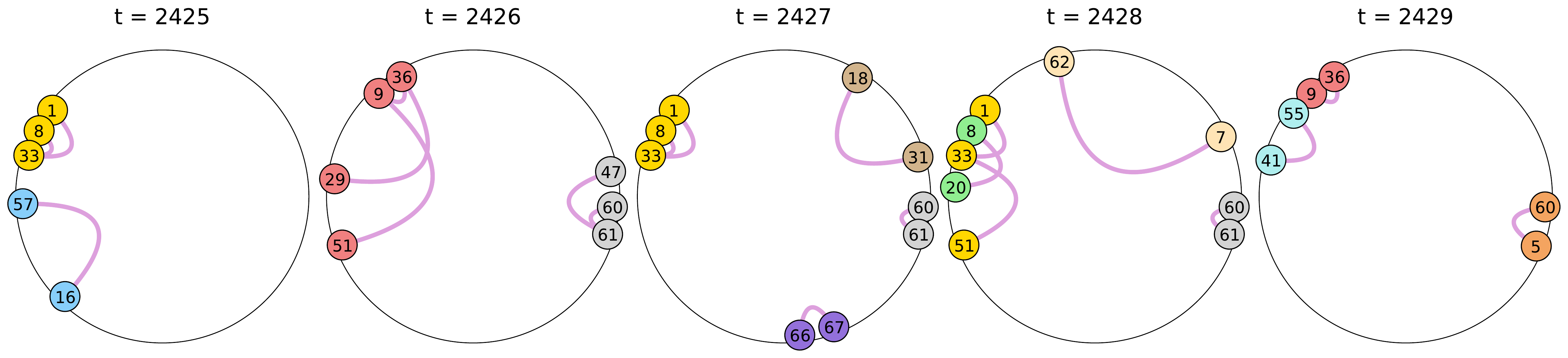}
\caption{Snapshots from the simulated counterpart of the hospital face-to-face interaction network generated by the dynamic-$\mathbb{S}^{1}$ (Sec.~\ref{sec:real_vs_modeled}). The snapshots correspond to time slots $t=2425$-$2429$. Each snapshot shows the interacting agents in their similarity space and the connections between them. The agents are colored according to the connected component where they belong, while the non-interacting agents in each snapshot, i.e., the agents with zero degree, are not shown to avoid clutter. The contact duration between agents $60$ and $61$ is three slots ($2426$-$2428$), while the inter-contact duration between agents $9$ and $36$ is two slots (2427, 2428). Agents $1$, $8$ and $33$ belong to a component forming both at $t=2425$ and $t=2427$ (recurrent component).
\label{fig:s1_snapshots}}
\end{figure*}

We note that the snapshots are conditionally independent given the agents' latent variables $\kappa_1, \theta_1, \ldots, \kappa_N, \theta_N$, but \emph{not} independent. In other words, even though each snapshot $G_t$ is constructed anew, there are correlations among the snapshots that are induced by the nodes' effective distances $\chi_{ij}$. In particular, nodes at smaller effective distances have higher chances of being connected in each snapshot, as dictated by the connection probability in~(\ref{eq:p_s1}). Fig.~\ref{fig:s1_snapshots} provides a visualization of snapshots generated by the model, where we see that agents at smaller similarity distances tend to stay connected in consecutive time slots and form recurrent components. We make the code implementing the model available at~\cite{modelURL}. Next, we compare the properties of synthetic networks generated by the model and real networks.

%%%%%%%%%%%%%%%%%%%%%%%%%%%%%%%%%%%%%%%%%%%%%%%%%%%%%%%%%%%%%%%%%%%%%%%
\section{Modeled vs. real networks}
\label{sec:real_vs_modeled}

\subsection{Overview of real networks}
\label{sec:real_nets}

We consider four face-to-face interaction networks from SocioPatterns~\cite{SocioPatterns}, which correspond to:  (i) a hospital ward in Lyon~\cite{HospitalData}; (ii) a primary school in Lyon~\cite{PrimaryData}; (iii) a high school in Marseilles~\cite{HighSchoolData}; and (iv) a scientific conference in Turin~\cite{ConferenceData}. These networks were captured over a period of $5$, $2$, $5$ and $2.5$ days, respectively. Each of their snapshots corresponds to a time slot of $20$ sec. We also consider the Bluetooth-based proximity network of the members of a residential community adjacent to a research university in North America, taken from the Friends and Family dataset~\cite{FFData}. The snapshots here correspond to slots of $5$ min, spanning the period October 2010 to May 2011. In all cases we number the slots and assign node IDs sequentially, $t=1,2,\ldots, \tau$ and $i=1,2, \ldots, N$. Table~\ref{tableReal} gives an overview of the data.
 
\begin{table}
\begin{tabular}{|c|c|c|c|c|c|c|c|c|}
\hline 
Network & $N$ & $\tau$ & $\bar{n}$  &  $\bar{d}$ & $\bar{k}_\textnormal{aggr}$\\ 
\hline 
Hospital & 75 & 17376 & 2.9 &  0.05 & 30\\ 
\hline 
Primary school & 242 & 5846 & 30 &  0.18 & 69\\ 
\hline 
High school & 327 & 18179 & 17 &  0.06 & 36\\ 
\hline 
Conference & 113 & 10618 & 3.3 &  0.03 & 39\\ 
\hline 
Friends \& Family & 131 & 57961 & 52 &  1.1 & 97\\
\hline 
\end{tabular}
\caption{Overview of the considered real networks. $N$ is the number of agents; $\tau$ is the total number of time slots;  $\bar{n}$ is the average number of interacting agents per slot; $\bar{d}$ is the average agent degree per slot; and $\bar{k}_\textnormal{aggr}$ is the average degree in the time-aggregated network (defined in Sec.~\ref{sec:properties}). Average values above $10$ have been rounded to the nearest integer.
\label{tableReal}}
\end{table}

\begin{figure}
\includegraphics[width=3.5in]{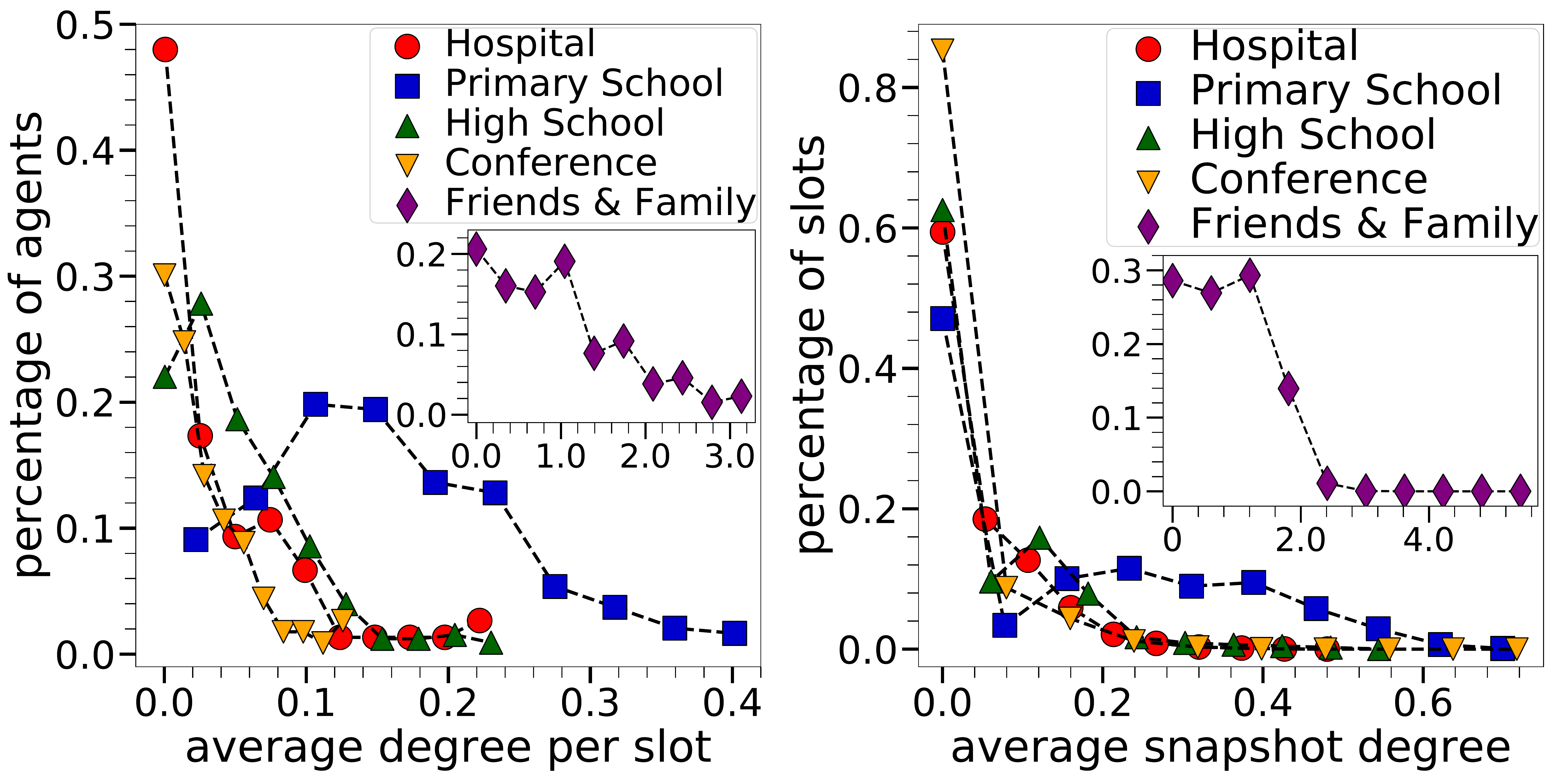}
\caption{Distribution of the average agent degree per slot (left) and of the average snapshot degree (right) in the considered networks. 
\label{figKappas}}
\end{figure}

We define the average degree per slot of agent $i$ as
\begin{align}
\label{eq:d_i_obs}
\bar{d}_i=\frac{1}{\tau}\sum_{t=1}^{\tau} d_{i, t},
\end{align}
where $d_{i,t} \geq 0$ is agent's $i$ degree in slot $t$, while the average agent (snapshot) degree in slot $t$ is
\begin{align}
\label{eq:k_t}
\bar{k}_t=\frac{1}{N}\sum_{i=1}^{N} d_{i, t}.
\end{align}
Fig.~\ref{figKappas} shows the distribution of $\bar{d}_i$ and $\bar{k}_t$ in the considered networks. The average agent degree per slot is
\begin{align}
\label{eq:k_obs}
\bar{d}=\frac{1}{N}\sum_{i=1}^{N} \bar{d}_i=\frac{1}{\tau} \sum_{t=1}^{\tau} \bar{k}_t.
\end{align}

%%%%%%%%%%%%%%%%%%%%%%%%%%%%%%%%%%%%%%%%%%%%%%%%%%%%%%%%%%%%%%%%%%%%%%%
\subsection{Modeled networks}
\label{sec:modeled_nets}

For each real network we construct its synthetic counterpart using the dynamic-$\mathbb{S}^{1}$. Each counterpart has the same number of nodes $N$ and duration $\tau$ as the corresponding real network, while the latent variable $\kappa_i$ of each agent $i=1, \ldots, N$ is set equal to the agent's average degree per slot in the real network,
\begin{align}
\label{eq:eq1}
\kappa_i=\bar{d}_i.
\end{align} 
Thus, the distribution of $\kappa_i$ is the corresponding empirical distribution in Fig.~\ref{figKappas} (left). The target average degree $\bar{k}_t$ in each snapshot $G_t$, $t=1, \ldots, \tau$, is set equal to the average degree in the corresponding real snapshot at slot $t$. Finally, the temperature $T$ is set such that the resulting average time-aggregated degree, $\bar{k}_\textnormal{aggr}$, is similar to the one in the real network---we analyze the dependence of $\bar{k}_\textnormal{aggr}$ on $T$ in Sec.~\ref{sec:barkaggr}.

In the counterparts, the expected degree of agent $i$ in slot $t$ is [Eq.~(\ref{eq:kappa})]
\begin{align}
\label{eq:kappa_t}
\bar{k}_t(\kappa_i)=\frac{\bar{k}_{t}}{\bar{d}}\kappa_i,
\end{align}
while agent's $i$ expected degree per slot is $\sum_{t=1}^{\tau}\bar{k}_t(\kappa_i)/\tau=\kappa_i$. The counterparts aim at capturing the variability in the number of interacting agents per slot since the probability that an agent $i$ interacts with at least one other agent in slot $t$ is
\begin{align}
I_{i,t}=1-\left[1-\frac{\bar{k}_t(\kappa_i)}{N-1}\right]^{N-1},
\end{align}
while  $\bar{k}_t(\kappa_i) \propto \bar{k}_{t} \kappa_i$.

%%%%%%%%%%%%%%%%%%%%%%%%%%%%%%%%%%%%%%%%%%%%%%%%%%%%%%%%%%%%%%%%%%%%%%%
\subsection{Properties of modeled vs. real networks}
\label{sec:properties}

\begin{figure*}
\includegraphics[width=3.5cm]{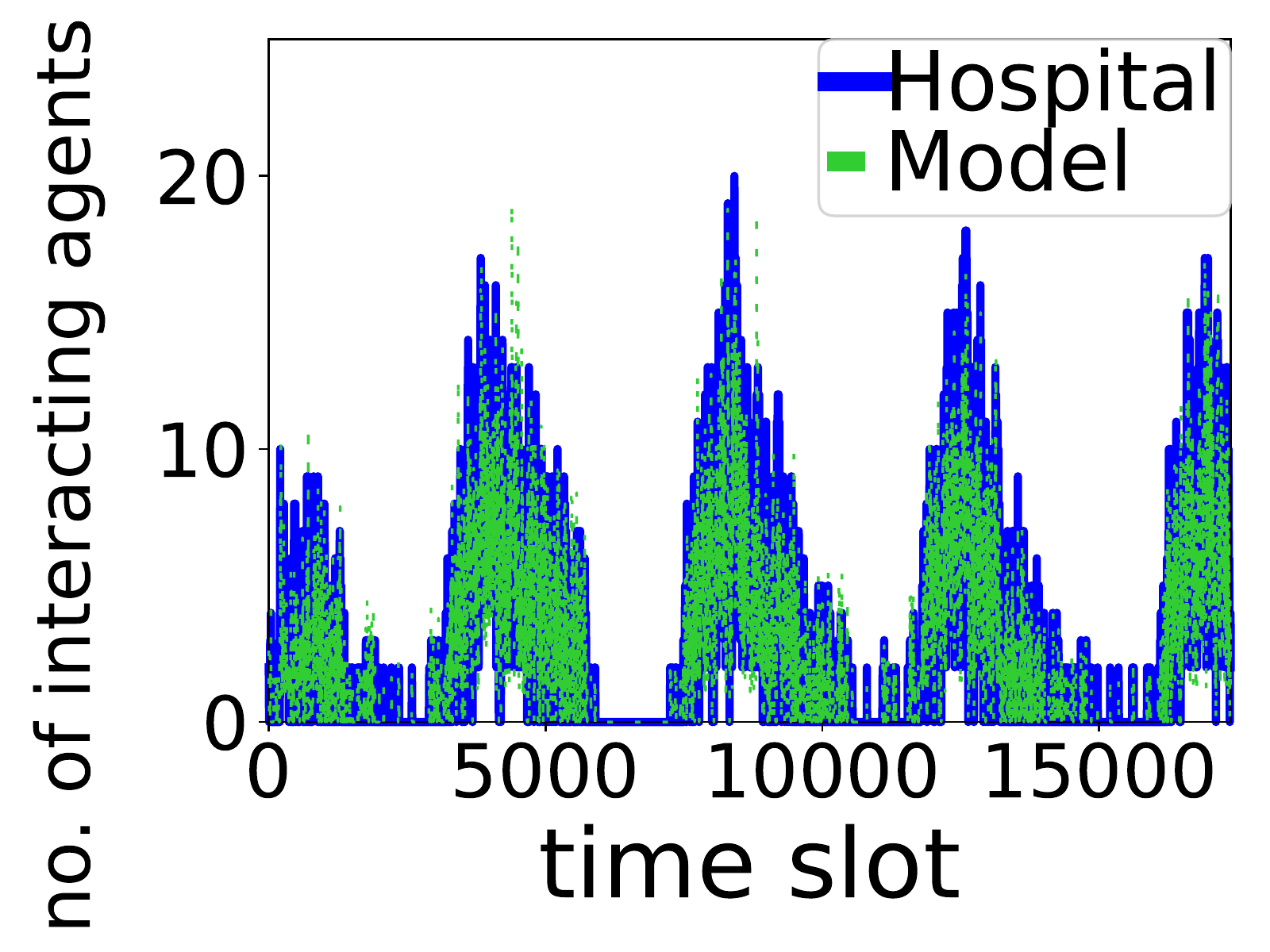}
\includegraphics[width=3.5cm]{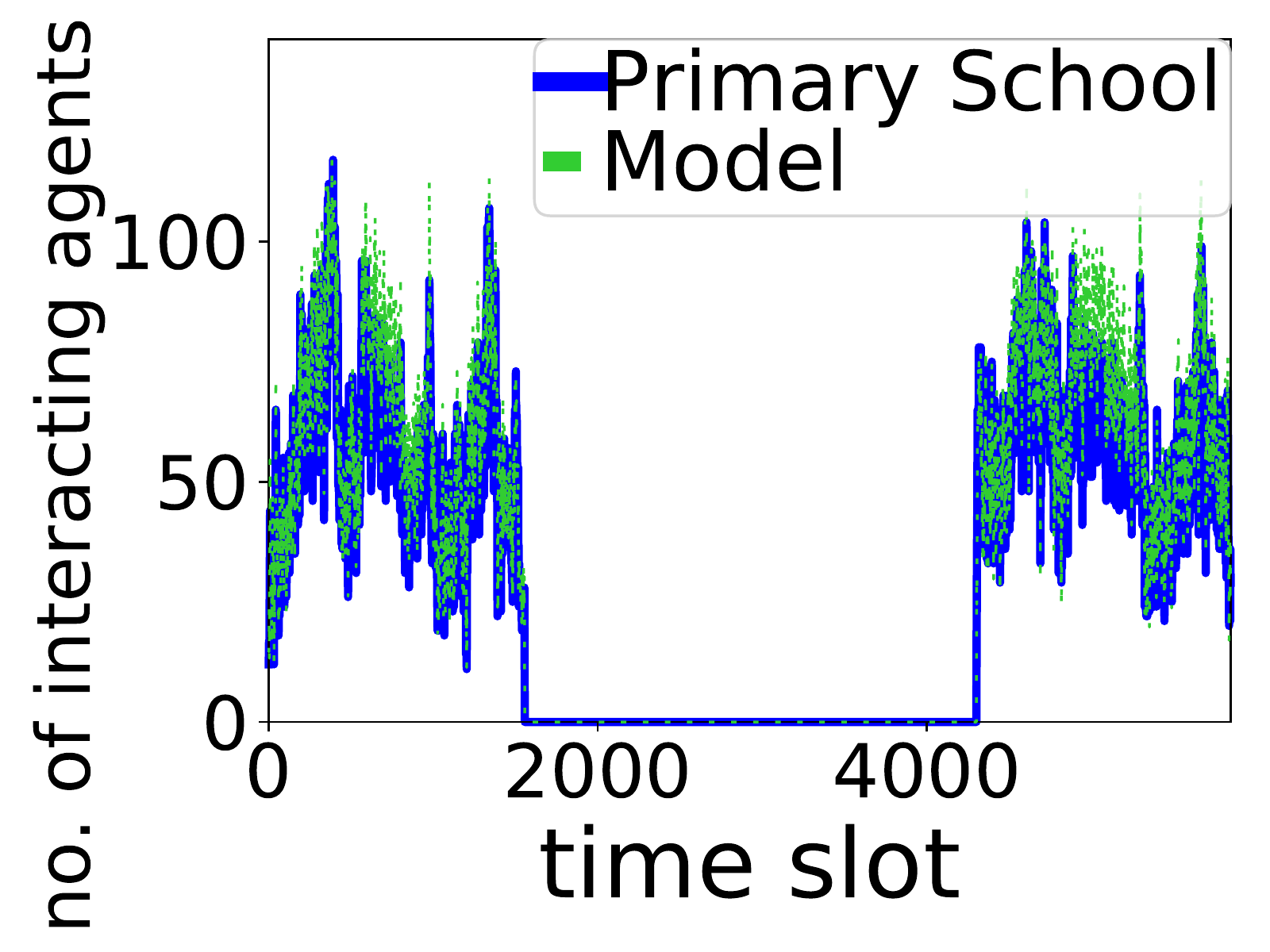}
\includegraphics[width=3.5cm]{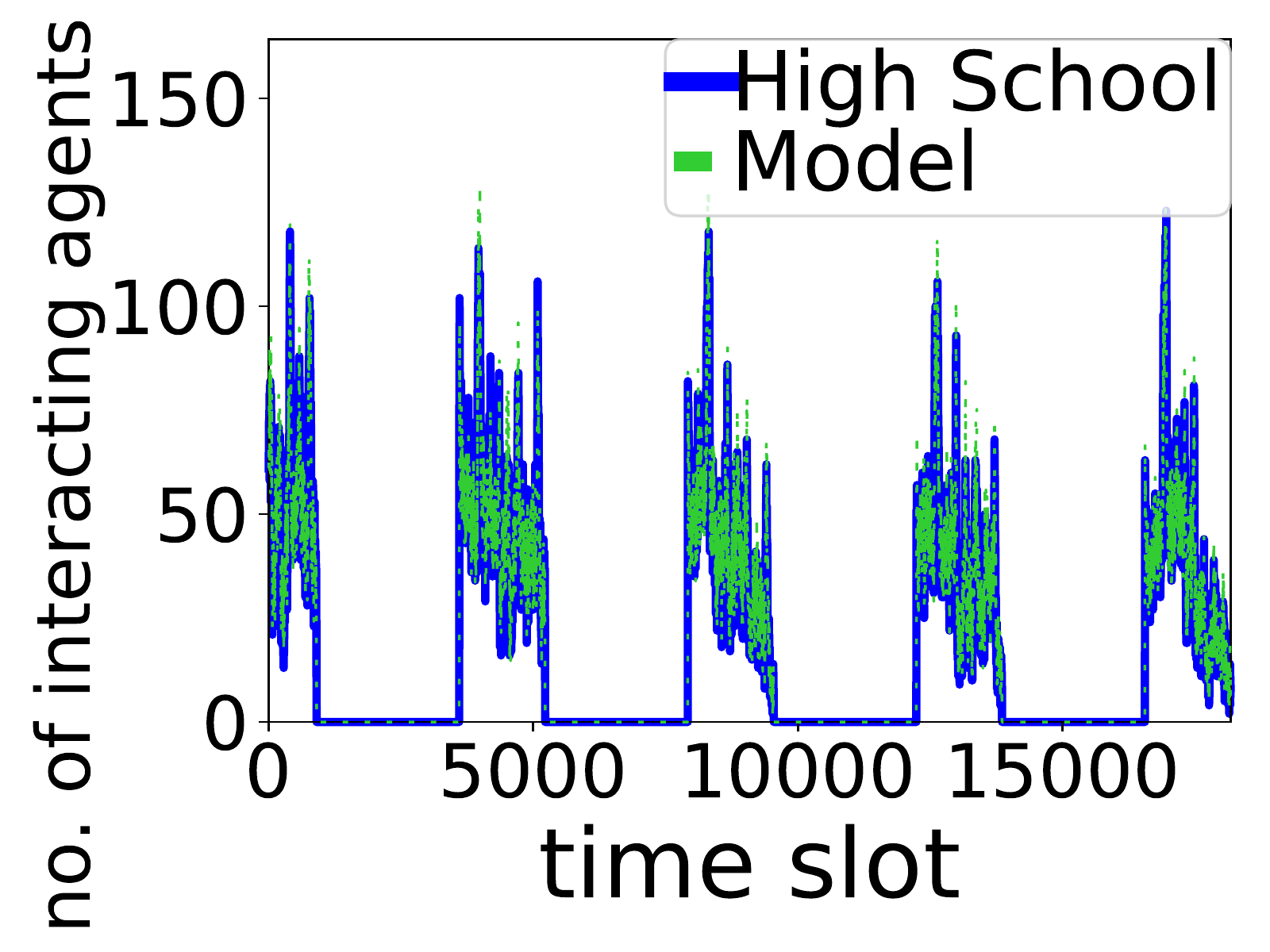}
\includegraphics[width=3.5cm]{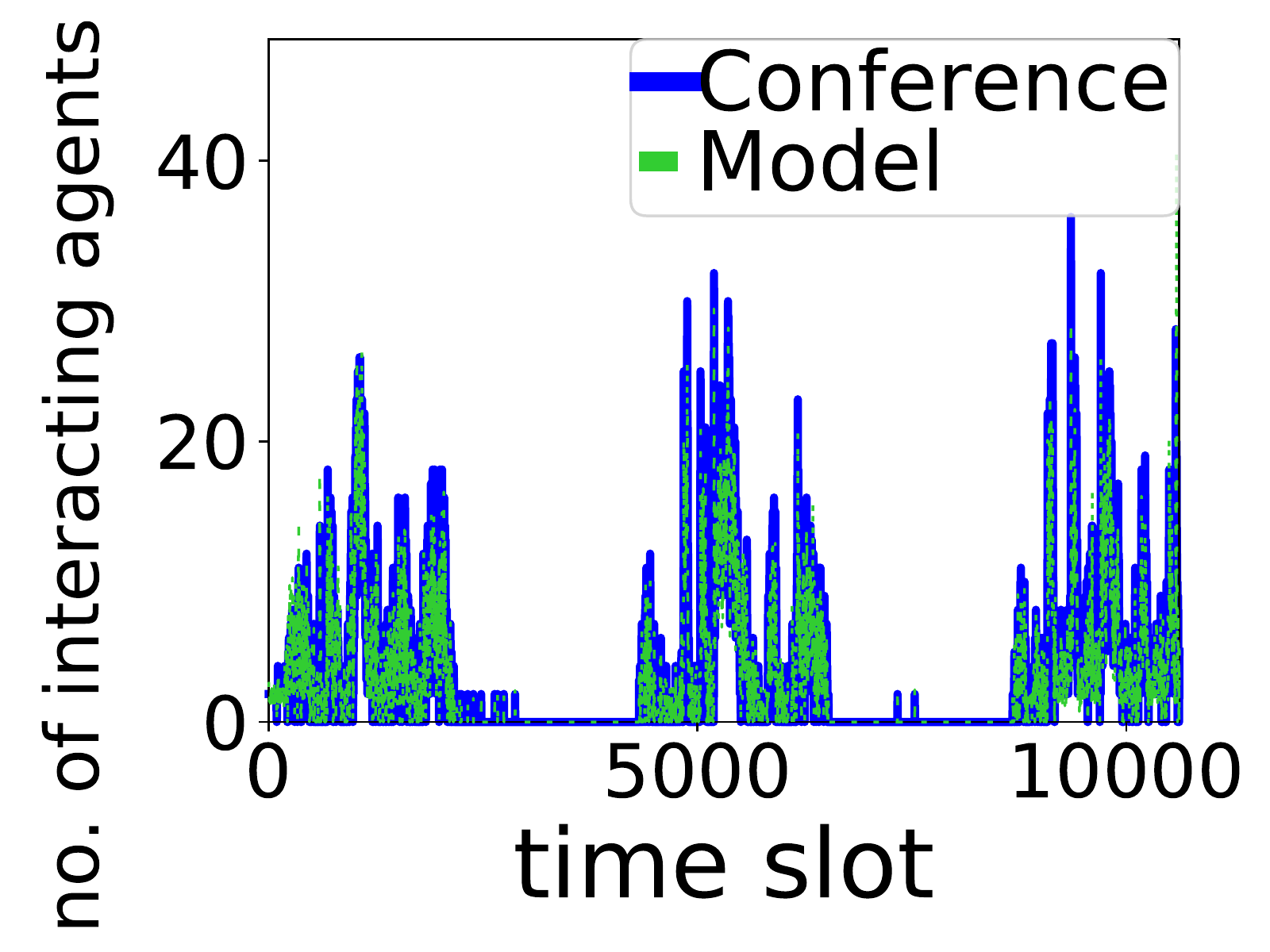}
\includegraphics[width=3.5cm]{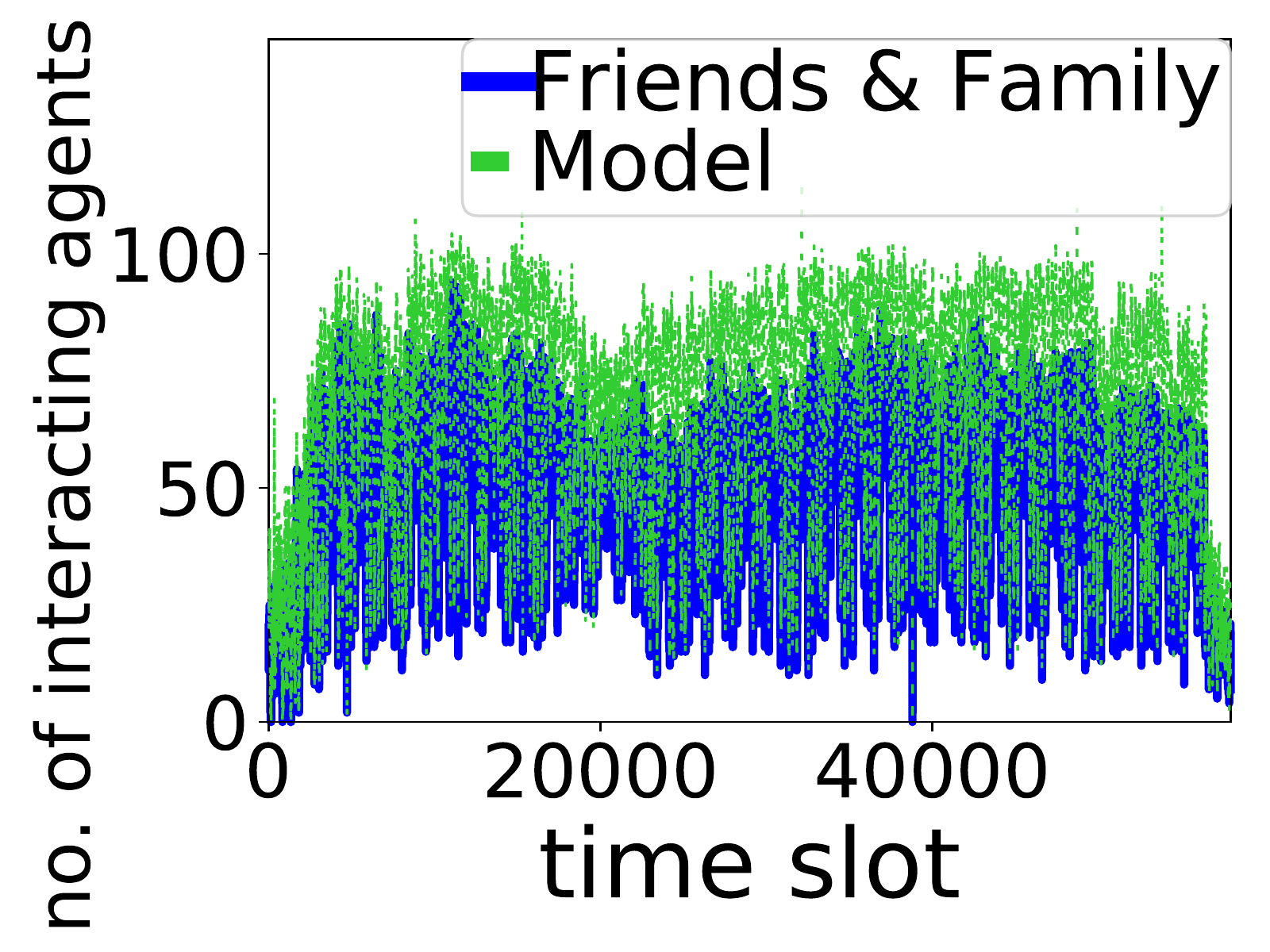}
\caption{Number of interacting agents per slot in real and modeled networks. In the first four plots the cycles of activity, i.e., the periods with high numbers of interacting agents, correspond to the consecutive observation days where the agents were present in the corresponding premises ($5$, $2$, $5$ and $2.5$ days, respectively.) There is a single activity cycle in the last plot, spanning the whole observation period---proximity in the Friends \& Family was constantly captured using mobile phones. 
\label{fignodes}}
\end{figure*}

Table~\ref{tableSimulated} gives an overview of the counterparts. We see that their characteristics are overall very similar to the ones of the real networks  (Table~\ref{tableReal}). Further, Fig.~\ref{fignodes} shows that the counterparts indeed capture the variability in the number of interacting agents per slot.

\begin{table}
\begin{tabular}{|c|c|c|c|c|c|c|c|c|}
\hline 
Modeled network & $N$ & $\tau$ & $\bar{n}$ & $\bar{d}$ & $\bar{k}_\textnormal{aggr}$ & $T$ \\ 
\hline 
Hospital & 75 & 17376 & 2.5 & 0.04 & 30 & 0.84 \\ 
\hline 
Primary school & 242 & 5846 & 33 & 0.17 & 69 & 0.72\\ 
\hline 
High school  & 327 & 18179 & 18 & 0.06 & 35 & 0.61\\ 
\hline 
Conference & 113 & 10618 & 2.9 & 0.03 & 30 & 0.85\\ 
\hline 
Friends \& Family & 131 & 57961 & 67 & 1.1 & 96 & 0.53\\
\hline 
\end{tabular}
\caption{Modeled counterparts. The values of $\bar{n}$, $\bar{d}$ and $\bar{k}_\textnormal{aggr}$ are averages over $20$ simulation runs except from the Friends \& Family where the averages are over $5$ runs. Average values above $10$ have been rounded to the nearest integer.
\label{tableSimulated}}
\end{table}

\begin{figure*}
\includegraphics[width=18cm, height=6cm]{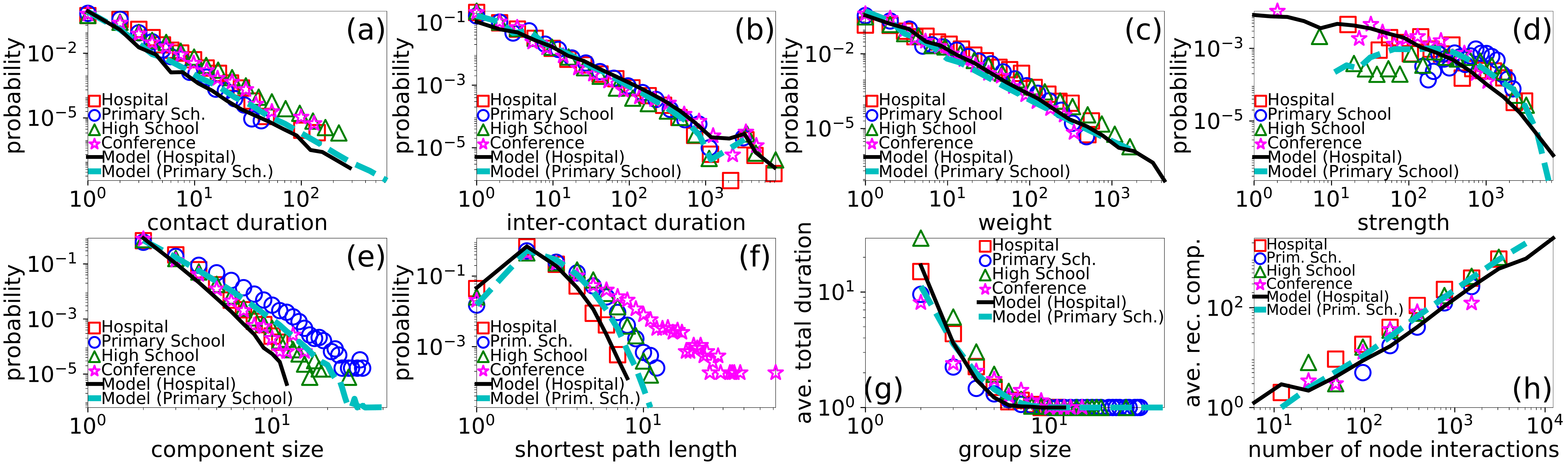}
\caption{Real face-to-face interaction networks vs. simulated networks with the dynamic-$\mathbb{S}^{1}$. 
\textbf{(a)}~Contact distribution. 
\textbf{(b)}~Inter-contact distribution. 
\textbf{(c)}~Weight distribution. 
\textbf{(d)}~Strength distribution. 
\textbf{(e)}~Distribution of component sizes. 
\textbf{(f)}~Distribution of shortest time-respecting path lengths.
\textbf{(g)}~Average total duration of a group as a function of its size. 
\textbf{(h)}~Average number of recurrent components where an agent participates as a function of the total number of interactions of the agent.
The results with the model are averages over $20$ simulation runs and correspond to the counterparts of the hospital and primary school. Similar results hold for the rest of the counterparts, not shown to avoid clutter. The probabilities in (a)-(f) represent relative frequencies, i.e., they are computed as $n_i/\sum_{j} n_j$, where $n_i$ is the number of samples that have value $i$. (a)-(d) have been binned logarithmically. Durations are measured in numbers of time slots.
\label{figAllProps}}
\end{figure*}

\begin{figure*}
\includegraphics[width=18cm, height=6cm]{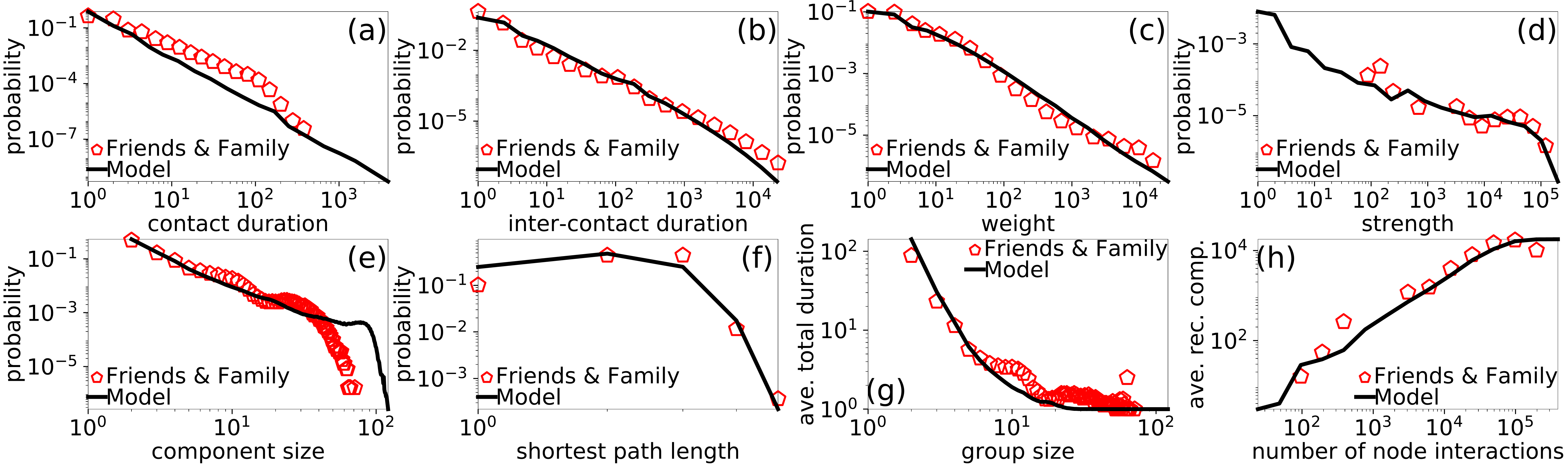}
\caption{Same as Fig.~\ref{figAllProps} but for the Friends \& Family proximity network and its modeled counterpart. The results with the model are averages over $5$ simulation runs.
\label{figFFProps}}
\end{figure*}

In Figs.~\ref{figAllProps} and~\ref{figFFProps} we compare a range of other properties between real and modeled networks, considered also in~\cite{Starnini2013, Starnini2016, flores2018}. These properties are:
\begin{itemize}[noitemsep]
\item[(a)] The aggregated contact distribution, i.e., the distribution of the number of slots that a pair of nodes remains connected.
\item[(b)] The aggregated inter-contact distribution, i.e., the distribution of the number of slots that a pair of nodes remains disconnected.
\item [(c)] The aggregated weight distribution, which is the distribution of the edge weights in the time-aggregated network. In this network two nodes are connected if they were connected in at least one slot, while the weight of an edge is the total number of slots that the two endpoints of the edge were connected. 
\item[(d)] The strength distribution, which is the distribution of the node strengths in the time-aggregated network. The strength of a node is the sum of the weights of all edges attached to the node. 
\item [(e)] The distribution of component sizes, which is the distribution of the number of nodes in the connected components formed throughout the observation period $\tau$.
\item[(f)] The distribution of the shortest time-respecting path lengths across all pairs of nodes. As an example, consider three nodes $i$, $k$ and $j$, where $i$ and $k$ connect at slot $t$ and $k$ and $j$ connect at slot $t' > t$. The time-respecting path between $i$ and $j$ is $i \to k \to j$ and has length $2$. The shortest time-respecting path between $i$ and $j$ is the shortest such path throughout the observation period. 
\item[(g)] The average total duration of a group as a function of its size. A group is a set of nodes forming a connected component. The total duration of a group is the total number of slots where the exact same set of nodes formed a connected component. For each group size we compute the average of this duration among groups with that specific size. 
\item[(h)] Finally, we consider the average number of recurrent components where an agent participates as a function of its total number of interactions (strength) throughout the observation period. A  connected component formed in a slot $t$ is called \emph{recurrent} if a connected component with exactly the same nodes was formed in a previous slot $t' < t$~\cite{flores2018}. We consider recurrent components consisting of at least three nodes.
\end{itemize}

Figs.~\ref{figAllProps} and~\ref{figFFProps} show that the dynamic-$\mathbb{S}^{1}$ reproduces all the above properties remarkably well.  A main exception are the longer paths in the conference~[Fig.~\ref{figAllProps}(f)], which can not be captured by the model. We also note that $\bar{k}_\textnormal{aggr}$ in conference's counterpart could not exceed $\approx 30$ (vs.~$39$ in the real network). Thus, the dynamic-$\mathbb{S}^{1}$ does not totally capture the characteristics of this network. Interestingly, this was also the case with the FDM~\cite{flores2018}. Finally, we note that the ability of the model to capture the properties of the considered networks is not due to mere calibration of expected node degrees. In Appendix~\ref{sec:scm}, we show that the configuration model~\cite{ChungLu2002,Newman2004} with the same calibration of expected node degrees, Eqs.~(\ref{eq:eq1},~\ref{eq:kappa_t}), cannot reproduce the abundance of recurrent components, nor the broad contact, inter-contact and weight distributions observed in the real systems. Further, in Sec.~\ref{sec:analysis} we prove these distributions in the dynamic-$\mathbb{S}^{1}$ and show that they do not depend on the distribution of the degree variables $\rho(\kappa)$. Below, we also investigate the \emph{pairwise} contact and inter-contact distributions in modeled and real networks.

%%%%%%%%%%%%%%%%%%%%%%%%%%%%%%%%%%%%%%%%%%%%%%%%%%%%%%%%%%%%%%%%%%%%%%%
\subsection{Pairwise contact and inter-contact distributions}
\label{sec:pairwise}

If the expected snapshot degrees, $\bar{k}_t, t=1, \ldots, \tau$, are independent and identically distributed, the pairwise contact and inter-contact distributions in the dynamic-$\mathbb{S}^{1}$ are geometric at $\tau \to \infty$~\footnote{For finite $\tau$ they are truncated geometric.}. Indeed, in this case the probability for two nodes $i, j$ with latent variables $\kappa_i, \kappa_j$ and angular distance $\Delta\theta_{ij}$ to remain connected for $t=1,2, \dots$ slots, is
\begin{align}
\label{eq:g1}
P_\textnormal{c}(t; \kappa_i, \kappa_j, \Delta\theta_{ij}) & =\bar{p}_{ij}^{t-1}  \left(1-\bar{p}_{ij}\right),\\
\nonumber \bar{p}_{ij} &\equiv \int p [\chi_{ij}(\bar{k})]f(\bar{k}) \mathrm{d} \bar{k},
\end{align}
where $p[\cdot]$ is the connection probability in Eq.~(\ref{eq:p_s1}), while $\chi_{ij}(\bar{k})$ is the effective distance between the two nodes, which depends on the average snapshot degree $\bar{k}$ [Eqs.~(\ref{eq:chi},~\ref{eq:mu})], whose PDF is denoted by $f(\cdot)$. Similarly, the probability that the two nodes remain disconnected for $t=1,2, \dots$ slots, is
\begin{align}
\label{eq:g2}
P_\textnormal{ic}(t; \kappa_i, \kappa_j, \Delta\theta_{ij}) =\left(1-\bar{p}_{ij}\right)^{t-1}\bar{p}_{ij}.
\end{align}
In general, these distributions are not geometric in the model as they depend on the stochastic process that describes the time evolution of the expected snapshot degrees. 

Previous studies have reported that a significant portion of pairwise inter-contact durations in real data can be fitted with exponential distributions~\cite{pairwise1, pairwise3}. Since the geometric distribution is the discrete analogue of the exponential distribution, these studies are in line with Eq.~(\ref{eq:g2}). Given these results, we check below how well the geometric distribution captures the pairwise contact and inter-contact distributions in the considered real systems and their modeled counterparts. 

For each pair of nodes we consider the sets of its contact and inter-contact durations in each of the activity cycles shown in Fig.~\ref{fignodes}. We consider sets with at least three distinct duration values. For each set we estimate the parameter of the geometric distribution, i.e., the success probability $p=1/m$, where $m$ is the mean of the durations in the set. Then, we draw the same number of samples as the number of durations in the set from a geometric distribution with parameter $p$. Subsequently, we use the two-sample Kolmogorov-Smirnov (KS) goodness of fit test~\cite{Massey2Samp,TaylorDGOF} to test the hypothesis that the values in the set and the sampled values have the same distribution. We recall that such a statistical test can only \emph{reject} or \emph{fail to reject} a given hypothesis for a given significance level $\alpha$. This level corresponds to the probability of incorrectly rejecting the hypothesis, while if the test fails to reject the hypothesis, we only know that this is true to a confidence level $1-\alpha$. We use $\alpha=0.01$, and find for each activity cycle the percentage of pairs for which the test failed to reject the hypothesis.  Table~\ref{tab:KStest} shows the average of this percentage across the activity cycles in each network, averaged across ten repetitions of the above procedure. The results for each counterpart are also averaged across ten different temporal network realizations.

\begin{table}
\centering
\small
\begin{tabular}{|c|c|c|c|}
\hline 
Network & \multicolumn{1}{|c|}{\textbf{Contact dist.}} & \multicolumn{2}{|c|}{\textbf{Inter-contact dist.}}\\
\cline{2-4}
& \textbf{geometric} & \textbf{geometric} & \textbf{log-normal} \\ 
\hline 
HP (model) & $98\%$  & $97\%$ & $99\%$ \\
HP (real) & $97\%$ & $69\%$ & $100\%$ \\
\hline
PS (model) & $100\%$ & $100\%$ & $99\%$ \\
PS (real) & $98\%$ & $69\%$ & $100\%$ \\
\hline
HS (model) & $98\%$  & $98\%$ & $98\%$ \\
HS (real) & $94\%$ & $65\%$ & $100\%$ \\
\hline
CF (model) & $95\%$ & $92\%$ & $99\%$ \\
CF (real) & $97\%$ & $64\%$ & $100\%$ \\
\hline
F \& F (model) & $80\% $ & $85\%$ & $87\%$ \\
F \& F (real) & $77\%$ & $60\%$ & $78\%$ \\
\hline
\end{tabular}
\caption{Percentage of pairs (rounded to the nearest integer) where the KS test failed to reject the hypothesis that their contact/inter-contact distribution is geometric. The table also shows the results where a log-normal distribution is assumed for the inter-contact durations; samples from the log-normal are rounded to the nearest integer before applying the KS test. (HP: Hospital; PS: Primary school; HS: High school; CF: Conference; F \& F: Friends and Family.)
\label{tab:KStest}}
\end{table}

We see in Table~\ref{tab:KStest} that the geometric distribution fits a high percentage of contact durations in both modeled and real networks. It also fits a high percentage of inter-contact durations in modeled networks, and a significant percentage of inter-contact durations in the real systems, which however is not as high as in the modeled networks. These results suggest that the model captures the variability of the contact durations in the real systems. However, it does not totally capture the variability of the inter-contact durations. 

\begin{figure}
\includegraphics[width=2.8in, height=1.7in]{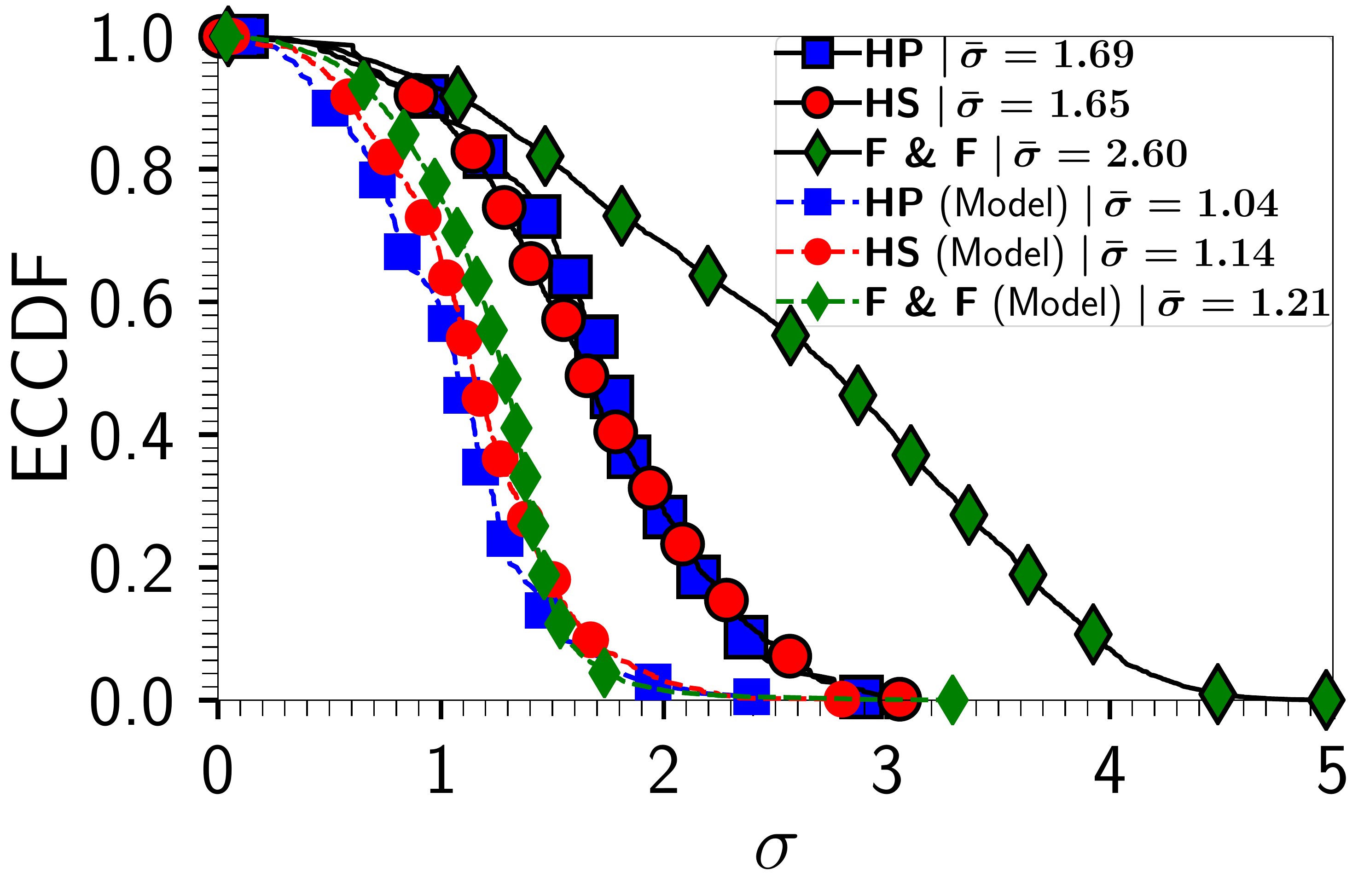}
\caption{Empirical complementary cumulative distribution function (ECCDF) of the estimated log-normal's $\sigma$ in real and modeled networks. The average ($\bar{\sigma}$) of each distribution is indicated in the legend. 
\label{fig:lognormalsigma}}
\end{figure}

To verify the last statement we also consider a log-normal distribution for the inter-contact durations, which offers a more versatile model to capture the variability in the distributions~\cite{pairwise1}. We recall that the PDF of the log-normal is $f(x)=1/(x \sigma \sqrt{2\pi}) e^{-(\ln{x}-\mu)^2/(2\sigma^2)}$, while its skewness is $(e^{\sigma^2}+2)\sqrt{e^{\sigma^2}-1}$. For each pair of nodes, the parameters $\mu$ and $\sigma^2$ are the mean and variance of the logarithms of its inter-contact durations. We see in Table~\ref{tab:KStest} that the log-normal better fits the inter-contact durations, especially in the real systems, as also observed in~\cite{pairwise1}. Further, Fig.~\ref{fig:lognormalsigma} shows that the inter-contact distributions in the real networks are indeed more skewed on average than in their counterparts. Nevertheless, the aggregated inter-contact distributions are very similar in real and synthetic systems  [Figs.~\ref{figAllProps}(b),~\ref{figFFProps}(b)]. In the next section we also see that paradigmatic dynamical processes perform similarly in the two. 

%%%%%%%%%%%%%%%%%%%%%%%%%%%%%%%%%%%%%%%%%%%%%%%%%%%%%%%%%%%%%%%%%%%%%%%
\section{Dynamical processes on modeled vs. real networks}
\label{sec:processes}

\begin{figure*}
\includegraphics[width=18cm]{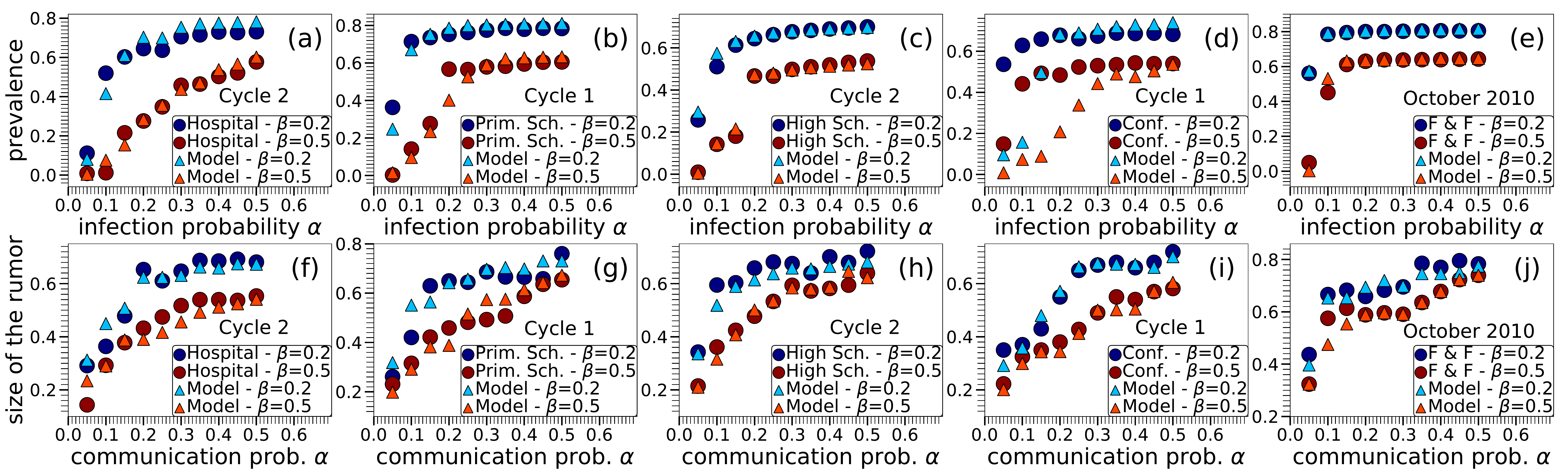}
\caption{Performance of the SIS and DK processes in real and modeled networks. Top row: prevalence of the SIS process as a function of the infection probability $\alpha$ for two recovery probabilities $\beta$. Bottom row: size of the rumor in the DK process as a function of the probability to communicate the rumor $\alpha$ for two stifling probabilities $\beta$. The results are averages over ten runs of each process in the activity cycles indicated in the plots. Each run of the SIS/DK process starts with a random set of infected/spreader agents that consists of $10\%$ of agents. The results for the modeled counterparts are also averaged across ten different temporal network realizations. 
\label{fig:dyn_proc}}
\end{figure*}

We consider the susceptible-infected-susceptible (SIS) epidemic spreading model~\cite{sis_ref} and the DK (Daley and Kendall) model for rumor spreading~\cite{Daley1965}. In the SIS each agent can be in one of two states, susceptible~(S) or infected~(I).  At any time slot an infected agent recovers with probability $\beta$ and becomes susceptible again, whereas infected agents infect the susceptible agents with whom they interact with probability $\alpha$. Thus, the transition of states is S~$\rightarrow$~I~$\rightarrow$~S. In the DK model each agent can be in one of three states, ignorant~(I), spreader~(S) or stifler~(R). An ignorant agent that interacts with a spreader receives the rumor with probability $\alpha$ and becomes a spreader, while a spreader that interacts with another spreader or a stifler becomes a stifler with probability $\beta$ and no longer communicates the rumor. The transition of states is I~$\rightarrow$~S~$\rightarrow$~R. 

To simulate the SIS process on temporal networks we use the dynamic SIS implementation of the Network Diffusion Library~\cite{NDlib}. We have also modified this library to implement the DK model. For the SIS process we consider the average percentage of infected agents per slot (prevalence), while for the DK process we consider the percentage of stiflers at the final slot (size of the rumor).  Fig.~\ref{fig:dyn_proc} shows that the two processes perform remarkably similar in real and modeled networks. The only exception is  in the performance of the SIS in the conference and its counterpart at low infection probabilities [Fig.~\ref{fig:dyn_proc}(d)]---a similar behavior has been observed in the FDM~\cite{flores2018} and it may be due to the fact that the models do not totally capture the characteristics of this network, as noted in Sec.~\ref{sec:properties}.

%%%%%%%%%%%%%%%%%%%%%%%%%%%%%%%%%%%%%%%%%%%%%%%%%%%%%%%%%%%%%%%%%%%%%%%
\section{Mathematical analysis}
\label{sec:analysis}

Here we perform a detailed mathematical analysis of the main properties of the dynamic-$\mathbb{S}^{1}$. To facilitate the analysis, we assume that the expected snapshot degree is the same in all time slots, $\bar{k}_t = \bar{k}$,~$\forall t$. This assumption renders the connection probability between two nodes [Eq.~(\ref{eq:p_s1})] the same in all slots. However, we illustrate that the analytical results match closely the simulation results from the modeled counterparts of real systems, where this assumption does not hold.

We show that for sparse snapshots, $\bar{k} \ll N$, and large durations $\tau$, the aggregated contact, inter-contact and weight distributions can be approximated by power laws with exponents $2+T$, $2-T$ and $1+T$, respectively, where $T \in (0,1)$ is the temperature in the connection probability.  Technically, we consider these distributions in the thermodynamic limit, $N \to \infty$, and show that they are power-laws with the aforementioned exponents at $\tau \to \infty$. Interestingly, these results do not depend on the distribution of the latent degree variables $\rho(\kappa)$. Further, we analyze the expected degree in the time-aggregated network, and show that in finite networks the expected strength of a node grows super-linearly with its time-aggregated degree, as empirically observed in prior studies~\cite{Starnini2013, StarniniDevices2017}. We begin with the contact distribution.

%%%%%%%%%%%%%%%%%%%%%%%%%%%%%%%%%%%%%%%%%%%%%%%%%%%%%%%%%%%%%%%%%%%%%%%
\subsection{Aggregated contact distribution}

The probability $r_\textnormal{c}(t; \kappa_i, \kappa_j, \Delta\theta_{ij})$ to observe a sequence of exactly $t=1, 2, \ldots, \tau-2$ consecutive slots where two nodes  $i, j$ with latent variables $\kappa_i, \kappa_j$ and angular distance $\Delta \theta_{ij}$ are connected, is the percentage of time $\tau$ where we observe a slot where these two nodes are not connected, followed by $t$ slots where they are connected, followed by a slot where they are not connected~\footnote{For brevity we ignore the cases where the first/last of the slots that two nodes can be connected starts/ends at the beginning/end of the observation period.}. For each duration $t$, there are $\tau-t-1$ possibilities where this duration can be realized. For instance, if $t=2$ the two nodes can be disconnected in slot $i-1$, connected in slots $i, i+1$, and disconnected in slot $i+2$, where $i = 2, \ldots, \tau-2$. Therefore, the percentage of observation time where a duration of $t$ slots can be realized is $(\tau-t-1)/\tau$. Since the two nodes are connected in each slot with probability $p(\chi_{ij})$ with $\chi_{ij}$ in Eq.~(\ref{eq:chi}), we have
\begin{align}
\label{eq:p_c}
r_\textnormal{c}(t; \kappa_i, \kappa_j, \Delta\theta_{ij})=\left(\frac{\tau-t-1}{\tau}\right)p(\chi_{ij})^{t} [1-p(\chi_{ij})]^2.
\end{align}
Removing the condition on $\Delta\theta_{ij}$, which is uniform on $[0, \pi]$, yields
\begin{align}
\nonumber r_\textnormal{c}(t; \kappa_i, \kappa_j ) =& \left(\frac{\tau-t-1}{\tau}\right) \frac{1}{\pi} \int \limits_0^\pi p(\chi_{ij})^{t} [1-p(\chi_{ij})]^2 \mathrm{d} \Delta\theta_{ij}\\
\nonumber=&  \left(\frac{\tau-t-1}{\tau}\right) \frac{2 \mu \kappa_i \kappa_j}{N}\\
\label{eq:full_contact_integral} 
\times& \int \limits_0^{\frac{N}{2\mu \kappa_i \kappa_j}} p(\chi_{ij})^{t} [1-p(\chi_{ij})]^2 \mathrm{d}\chi_{ij}\\
\nonumber =&  \left(\frac{\tau-t-1}{\tau}\right) \left(\frac{N}{2 \mu \kappa_i \kappa_j}\right)^{2/T} \left(\frac{T}{2+T}\right)\\
\nonumber \times& {}_2 F_{1}\left[t+2, 2+T, 3+T, -\left(\frac{N}{2 \mu \kappa_i \kappa_j}\right)^{1/T}\right],
\end{align}
where ${}_2 F_1[a, b, c; z]$ is the Gauss hypergeometric function~\cite{special_functions_book}.  At $N \to \infty$, the integral in~(\ref{eq:full_contact_integral}) simplifies for $T \in (0, 1)$ and $t \geq 1$, to
\begin{align}
\label{eq:int1}
 \int \limits_0^{\infty} p(\chi_{ij})^{t} [1-p(\chi_{ij})]^2 \mathrm{d}\chi_{ij}=\frac{T \Gamma{(2+T)}\Gamma{(t-T)}}{\Gamma{(t +2)}},
\end{align}
where $\Gamma(z)$ is the complete gamma function, $\Gamma(z)=\int_{0}^{\infty} x^{z-1} e^{-x}  \mathrm{d} x$, $z > 0$~\footnote{If $z$ is a positive integer then $\Gamma{(z)}=(z-1)!$.}. From~(\ref{eq:full_contact_integral}, \ref{eq:int1}), we have
\begin{align}
\label{eq:lim_1}
\nonumber N r_\textnormal{c}(t; \kappa_i, \kappa_j ) \xrightarrow{ N \to \infty }& \left(\frac{\tau-t-1}{\tau}\right)2 \mu \kappa_i \kappa_j \\
\times& \frac{T \Gamma{(2+T)}\Gamma{(t-T)}}{\Gamma{(t +2)}}.
\end{align}
Removing the condition on $\kappa_i$ and $\kappa_j$, gives
\begin{align}
\label{eq:lim_1_uncon}
\nonumber N r_\textnormal{c}(t) &= N \int \int r_c(t; \kappa_i, \kappa_j) \rho(\kappa_i) \rho(\kappa_j) \mathrm{d} \kappa_i \mathrm{d} \kappa_j\\
& \xrightarrow{ N \to \infty } \left(\frac{\tau-t-1}{\tau}\right) \frac{2 \mu \bar{\kappa}^2 T \Gamma{(2+T)}\Gamma{(t-T)}}{\Gamma{(t +2)}}.
\end{align}

\begin{figure}
\includegraphics[width=2.8in]{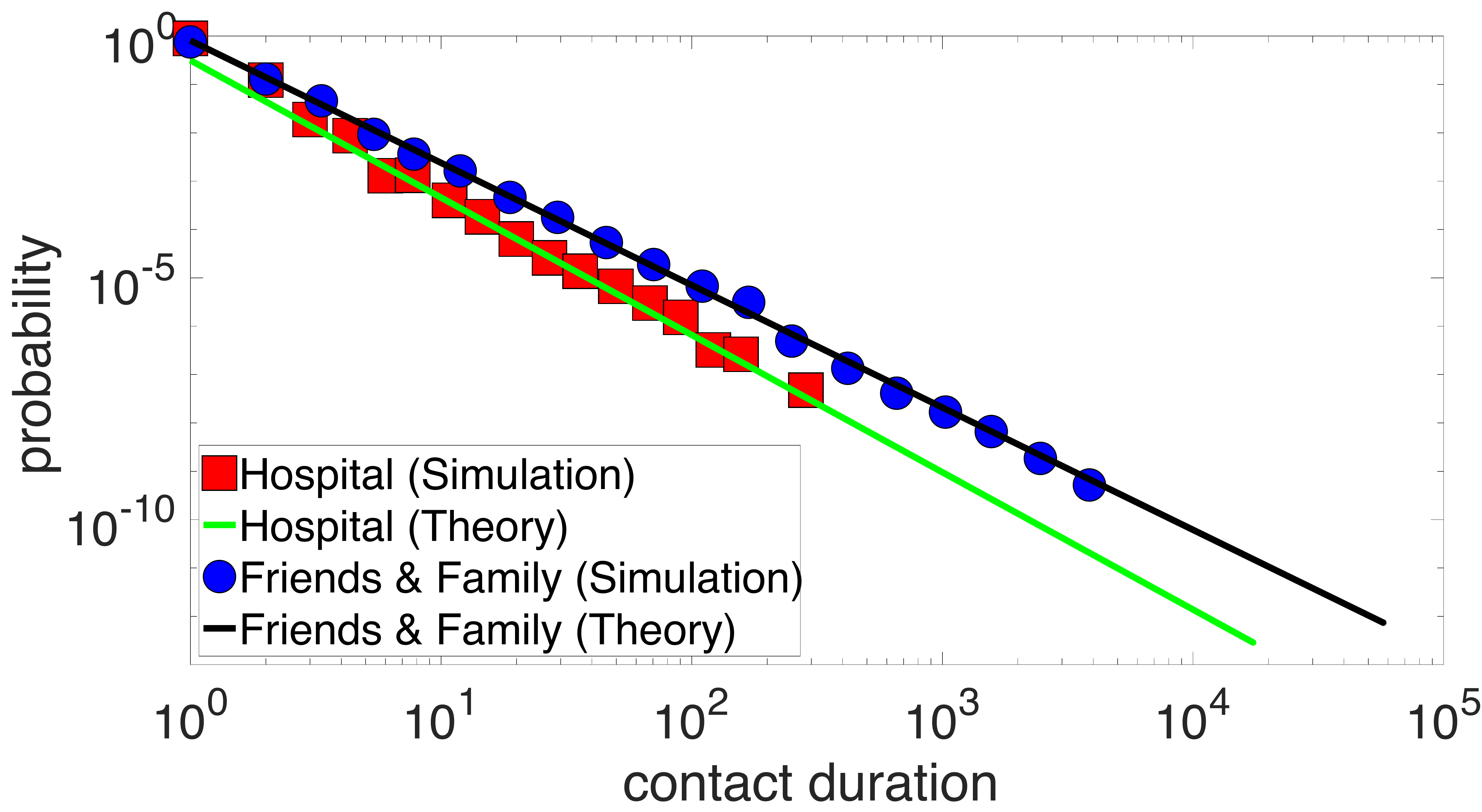}
\caption{Aggregated contact distribution in the simulated counterparts of the hospital and Friends \& Family (Sec.~\ref{sec:modeled_nets})~vs.~theoretical prediction in~(\ref{eq:p_c_final_limit}) with $T=0.84, 0.53$. Similar results hold for the rest of the counterparts.
\label{fig:validation1}}
\end{figure}

The aggregated contact distribution, $P_{\textnormal{c}}(t)$, is the probability that two nodes are connected for exactly $t$ consecutive slots  given that $t \geq 1$,
\begin{align}
\label{eq:p_c_norm}
P_{\textnormal{c}}(t) =\frac{r_\textnormal{c}(t)}{\sum_{t=1}^{\tau-2} r_\textnormal{c}(t)}.
\end{align}
From~(\ref{eq:lim_1_uncon}, \ref{eq:p_c_norm}), we have
\begin{align}
\label{eq:p_c_final}
P_{\textnormal{c}}(t) \xrightarrow{ N \to \infty } \frac{(\tau-t-1)}{g(\tau)} \frac{\Gamma{(t-T)}}{\Gamma{(t +2)}} \approx \frac{(\tau-t-1)}{g(\tau)} \frac{1}{t^{2+T}},
\end{align}
where
\begin{align}
\nonumber g(\tau) \equiv \frac{\left[(\tau-1)T-1\right]\Gamma{(1-T)}}{T+T^2}+\frac{\Gamma{(\tau-T)}}{(T+T^2)\Gamma{(\tau)}}.
\end{align}
The approximation in~(\ref{eq:p_c_final}) uses the facts $\Gamma{(t - T)} \approx t^{-T}\Gamma{(t)}$ and $\Gamma{(t+2)} \approx t^2 \Gamma{(t)}$, which hold for $t \gg 1$. We see from~(\ref{eq:p_c_final}) that for $t \ll \tau$, $P_{\textnormal{c}}(t)$ is approximately a power law with exponent $2+T$. At $\tau \to \infty$, we have a pure power law
\begin{align}
\label{eq:p_c_final_limit}
P_{\textnormal{c}}(t) \xrightarrow{\substack{N \to \infty \\ \tau \to \infty}}  \frac{1+T}{\Gamma{(1-T})}\frac{\Gamma{(t-T)}}{\Gamma{(t +2)}} \approx \frac{1+T}{\Gamma{(1-T})} \frac{1}{t^{2+T}}.
\end{align}
Fig.~\ref{fig:validation1} shows that~(\ref{eq:p_c_final_limit}) provides an excellent approximation to simulation results.

From~(\ref{eq:p_c_final}), the expected contact duration in the thermodynamic limit is
\begin{align}
\nonumber \bar{t}_{\textnormal{c}} \xrightarrow{ N \to \infty } & \sum_{t=1}^{\tau-2} t \frac{(\tau-t-1)}{g(\tau)} \frac{\Gamma{(t-T)}}{\Gamma{(t +2)}}\\
=&\frac{\Gamma{(2-T)}\Gamma{(\tau+1)}-\Gamma{(\tau-T)}[(1+T)\tau-2T]}{\Gamma{(2-T)} [(\tau-1)T-1]\Gamma{(\tau)}+\Gamma{(\tau-T)}(1-T)}.
\end{align}
At $\tau \to \infty$, the last relation simplifies to
\begin{align}
\bar{t}_{\textnormal{c}} \xrightarrow{\substack{N \to \infty \\ \tau \to \infty}}  \frac{1}{T}. 
\end{align}
Next, we derive the aggregated inter-contact distribution following the same steps. 

%%%%%%%%%%%%%%%%%%%%%%%%%%%%%%%%%%%%%%%%%%%%%%%%%%%%%%%%%%%%%%%%%%%%
\subsection{Aggregated inter-contact distribution}

Let $r_\textnormal{ic}(t; \kappa_i, \kappa_j, \Delta\theta_{ij})$ be the probability to observe a slot where two nodes  $i, j$ with latent variables $\kappa_i, \kappa_j$ and angular distance $\Delta \theta_{ij}$ are connected, followed by $t$ slots where they are not connected, followed by a slot where they are again connected. We have
\begin{align}
\label{eq:p_i}
r_\textnormal{ic}(t; \kappa_i, \kappa_j, \Delta\theta_{ij})=\left(\frac{\tau-t-1}{\tau}\right)p(\chi_{ij})^{2} [1-p(\chi_{ij})]^t.
\end{align}
Removing the condition on $\Delta \theta_{ij}$, yields
\begin{align}
\nonumber r_\textnormal{ic}(t; \kappa_i, \kappa_j ) =&  \left(\frac{\tau-t-1}{\tau}\right) \frac{1}{\pi} \int \limits_0^\pi p(\chi_{ij})^{2} [1-p(\chi_{ij})]^t \mathrm{d} \Delta\theta_{ij}\\
\nonumber =&  \left(\frac{\tau-t-1}{\tau}\right) \frac{2 \mu \kappa_i \kappa_j}{N}\\
\label{eq:full_intercontact_integral}
\times & \int \limits_0^{\frac{N}{2\mu \kappa_i \kappa_j}} p(\chi_{ij})^{2} [1-p(\chi_{ij})]^t \mathrm{d}\chi_{ij}\\
\nonumber =&  \left(\frac{\tau-t-1}{\tau}\right) \left(\frac{N}{2 \mu \kappa_i \kappa_j}\right)^{t/T} \left(\frac{T}{t+T}\right)\\
\nonumber \times& {}_2 F_{1}\left[t+T, t+2, t+T+1, -\left(\frac{N}{2 \mu \kappa_i \kappa_j}\right)^{1/T}\right].
\end{align}
At $N \to \infty$, the integral in~(\ref{eq:full_intercontact_integral}) simplifies for $T \in (0, 1)$, to
\begin{align}
\label{eq:int2}
\int \limits_0^{\infty} p(\chi_{ij})^{2} [1-p(\chi_{ij})]^t \mathrm{d}\chi_{ij}=\frac{T \Gamma{(2-T)} \Gamma{(t+T)}}{\Gamma{(t+2)}}.
\end{align}
From~(\ref{eq:full_intercontact_integral}, \ref{eq:int2}), and after removing the condition on $\kappa_i$ and $\kappa_j$, we have 
\begin{align}
\label{eq:lim_2_uncon}
N r_\textnormal{ic}(t) \xrightarrow{ N \to \infty } \left(\frac{\tau-t-1}{\tau}\right) \frac{2 \mu \bar{\kappa}^2 T \Gamma{(2-T)} \Gamma{(t+T)}}{\Gamma{(t+2)}}.
\end{align}

\begin{figure}
\includegraphics[width=2.8in]{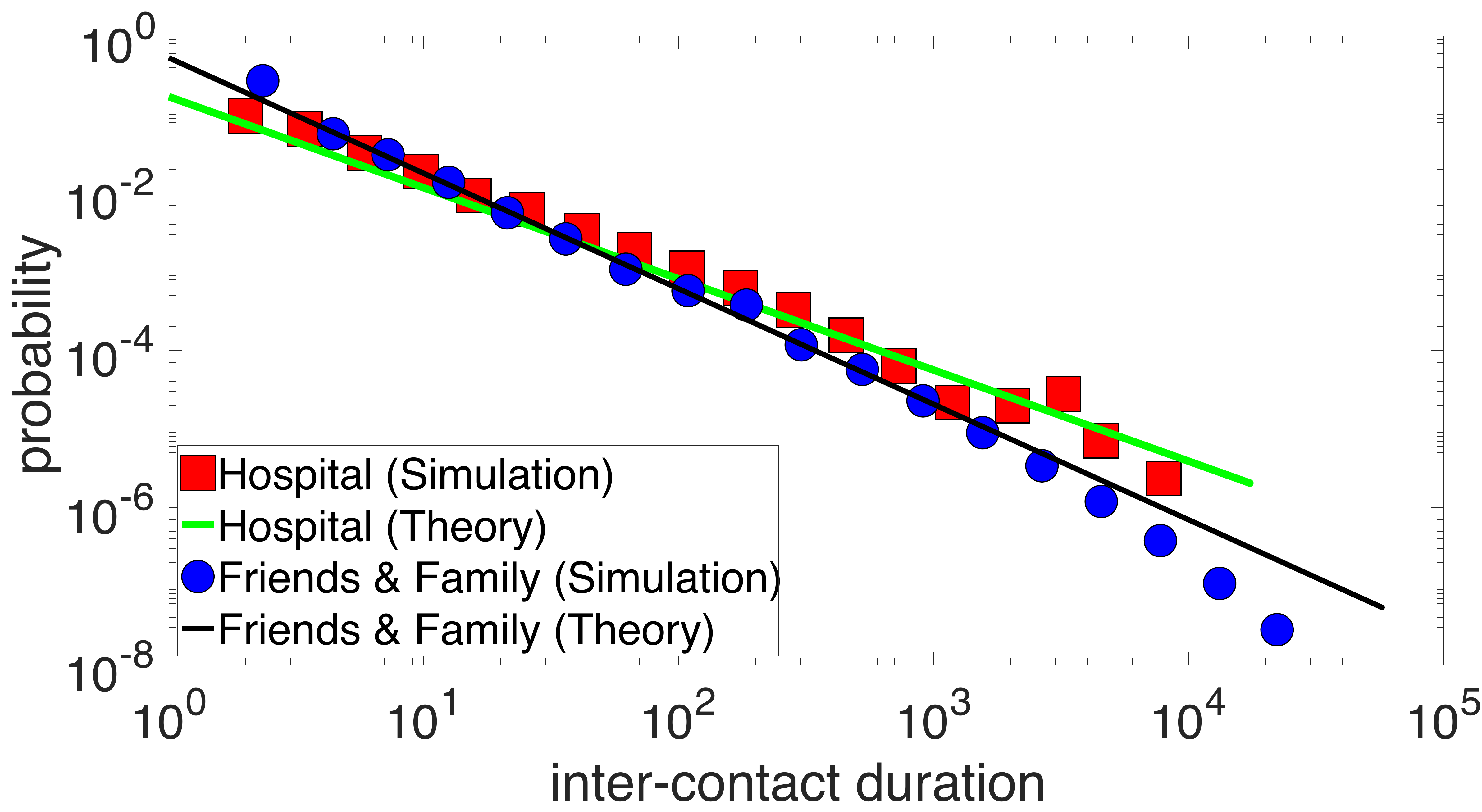}
\caption{Aggregated inter-contact distribution in the simulated counterparts of the hospital and Friends \& Family (Sec.~\ref{sec:modeled_nets})~vs.~theoretical prediction in~(\ref{eq:p_ic_final_limit}) with $T=0.84, 0.53$. 
Similar results hold for the rest of the counterparts.
\label{fig:validation2}}
\end{figure}

The aggregated inter-contact distribution, $P_{\textnormal{ic}} (t)$, is the probability that two nodes are disconnected for exactly $t$ consecutive slots given that $t \geq 1$,
\begin{align}
\label{eq:p_ic_norm}
P_{\textnormal{ic}}(t) =\frac{r_\textnormal{ic}(t)}{\sum_{t=1}^{\tau-2} r_\textnormal{ic}(t)}.
\end{align}
From~(\ref{eq:lim_2_uncon}, \ref{eq:p_ic_norm}), we have
\begin{align}
\label{eq:p_ic_final}
P_{\textnormal{ic}}(t) \xrightarrow{ N \to \infty } \frac{(\tau-t-1)}{h(\tau)} \frac{\Gamma{(t+T)}}{\Gamma{(t+2)}}\approx \frac{(\tau-t-1)}{h(\tau)} \frac{1}{t^{2-T}},
\end{align}
where
\begin{align}
\nonumber h(\tau) \equiv \frac{[(\tau-1)T+1]\Gamma{(1+T)}}{T-T^2}-\frac{\Gamma{(\tau+T)}}{(T-T^2)\Gamma{(\tau)}}. 
\end{align}
The approximation in~(\ref{eq:p_ic_final}) holds for $t \gg 1$. For $t \ll \tau$, $P_{\textnormal{ic}}(t)$ is approximately a power law with exponent $2-T$. At $\tau \to \infty$, we have a pure power law
\begin{align}
\label{eq:p_ic_final_limit}
P_{\textnormal{ic}}(t) \xrightarrow{\substack{N \to \infty \\ \tau \to \infty}} & \frac{1-T}{\Gamma{(1+T})}\frac{\Gamma{(t+T)}}{\Gamma{(t+2)}}\approx \frac{1-T}{\Gamma{(1+T})} \frac{1}{t^{2-T}}.
\end{align}
Fig.~\ref{fig:validation2} juxtaposes~(\ref{eq:p_ic_final_limit}) against simulation results. 

From~(\ref{eq:p_ic_final}), the expected inter-contact duration in the thermodynamic limit is
\begin{align}
\nonumber \bar{t}_{\textnormal{ic}} \xrightarrow{ N \to \infty } & \sum_{t=1}^{\tau-2} t \frac{(\tau-t-1)}{h(\tau)} \frac{\Gamma{(t+T)}}{\Gamma{(t+2)}}\\
=&\frac{\Gamma{(\tau+T)[(1-T)\tau+2T]}-\Gamma{(2+T)} \Gamma{(\tau+1)}}{\Gamma{(2+T)}[(\tau-1)T+1]\Gamma{(\tau)}-\Gamma{(\tau+T)}(1+T)}.
\end{align}
The above relation increases approximately exponentially with $T \in (0,1)$, and diverges at $\tau \to \infty$, 
\begin{align}
\bar{t}_{\textnormal{ic}} \xrightarrow{\substack{N \to \infty \\ \tau \to \infty}} \infty.
\end{align}
We proceed with the weight distribution.

%%%%%%%%%%%%%%%%%%%%%%%%%%%%%%%%%%%%%%%%%%%%%%%%%%%%%%%%%%%%%%%%%%%%
\subsection{Aggregated weight distribution}

The probability that two nodes $i, j$ with latent variables $\kappa_i, \kappa_j$ and angular distance $\Delta \theta_{ij}$ are connected in $t=0,1,\ldots, \tau$ slots, is given by the binomial distribution
\begin{align}
\label{eq:weights_binomial}
r_\textnormal{w}(t; \kappa_i, \kappa_j, \Delta\theta_{ij})= {\tau \choose t} p(\chi_{ij})^{t}[1-p(\chi_{ij})]^{\tau-t}.
\end{align}
Removing the condition on $\Delta\theta_{ij}$, yields
\begin{align}
\nonumber r_\textnormal{w}(t; \kappa_i, \kappa_j) = \frac{1}{\pi} {\tau \choose t} \int \limits_0^\pi p(\chi_{ij})^{t}[1-p(\chi_{ij})]^{\tau-t}  \mathrm{d} \Delta\theta_{ij}\\
\nonumber =\frac{2 \mu \kappa_i \kappa_j T}{N}  {\tau \choose t} \int \limits_{u_{ij}^\textnormal{min}}^{1} u_{ij}^{t-T-1}(1-u_{ij})^{\tau-t+T-1}  \mathrm{d} u_{ij}\\
\label{eq:uncon_w}
\nonumber = \frac{2 \mu \kappa_i \kappa_j}{N} \frac{T \Gamma{(\tau+1)}}{\Gamma{(\tau-t+1)}\Gamma{(t+1)}} \Bigg [\frac{\Gamma{(\tau-t+T)}\Gamma{(t-T)}}{\Gamma{(\tau)}}\\
-\frac{{(u_{ij}^\textnormal{min})}^{t-T}}{t-T} {}_2 F_{1}(t-T, 1-\tau-T+t, t-T+1, u_{ij}^\textnormal{min})\Bigg],
\end{align}
where
\begin{align} 
\label{eq:u_min}
u_{ij}^\textnormal{min} \equiv \frac{1}{1+\left(\frac{N}{2\mu \kappa_i \kappa_j}\right)^{1/T}}.
\end{align}
To reach~(\ref{eq:uncon_w}), we perform the change of integration variable $u_{ij} \equiv p(\chi_{ij})$ and express the binomial coefficient in terms of gamma functions, ${\tau \choose t} = \Gamma{(\tau+1)}/[\Gamma{(\tau-t+1)} \Gamma{(t+1)}]$. 

At $N \to \infty$, $u_{ij}^\textnormal{min} \to 0$, and the second term inside the brackets in~(\ref{eq:uncon_w}) vanishes for $T \in (0, 1)$ and $t \geq 1$. Removing the condition on $\kappa_i$ and $\kappa_j$, we have
\begin{align}
\label{eq:lim_3_uncon}
N r_\textnormal{w}(t)  \xrightarrow{ N \to \infty } \frac{2\mu \bar{\kappa}^2 T \tau \Gamma{(\tau-t+T)}\Gamma{(t-T)}}{\Gamma{(\tau-t+1)}\Gamma{(t+1)}}.
\end{align}
For $t=0$, we can write
\begin{align}
\label{eq:lim_4_uncon}
N[1-r_\textnormal{w}(0)]= N\sum_{t=1}^{\tau} r_\textnormal{w}(t)  \xrightarrow{ N \to \infty }  \frac{2\mu \bar{\kappa}^2 \Gamma{(1-T)}\Gamma{(\tau+T)}}{\Gamma{(\tau)}}.
\end{align}

The aggregated weight distribution, $P_{\textnormal{w}}(t)$, is the probability that two nodes are connected in $t$ slots given that $t \geq 1$,
\begin{align}
\label{eq:p_w_norm}
 P_{\textnormal{w}}(t) =\frac{r_\textnormal{w}(t)}{\sum_{t=1}^{\tau} r_\textnormal{w}(t)}.
\end{align}
From~(\ref{eq:lim_3_uncon}, \ref{eq:p_w_norm}), we have
\begin{align}
\label{eq:p_w_final}
 P_{\textnormal{w}}(t) \xrightarrow{ N \to \infty } &  \frac{1}{w(\tau)} \frac{\Gamma{(\tau-t+T)}\Gamma{(t-T)}}{\Gamma{(\tau-t+1)}\Gamma{(t+1)}}\\
 \label{eq:p_w_final_approx}
 \approx& \frac{1}{w(\tau) (\tau-t)^{1-T}} \frac{1}{t^{1+T}},
\end{align}
where
\begin{align}
\nonumber w(\tau) \equiv \frac{\Gamma{(1-T)}\Gamma{(\tau+T)}}{T \Gamma{(\tau+1)}}. 
\end{align}
The approximation in~(\ref{eq:p_w_final_approx}) holds for $1 \ll t \ll \tau$. We see from~(\ref{eq:p_w_final_approx}) that for $t \ll \tau$, $P_{\textnormal{w}}(t)$ is approximately a power law with exponent $1+T$. At $\tau \to \infty$, we have a pure power law
\begin{align}
\label{eq:p_w_final_limit1}
P_{\textnormal{w}}(t) \xrightarrow{\substack{N \to \infty \\ \tau \to \infty}}  \frac{T}{\Gamma{(1-T)}} \frac{\Gamma{(t-T)}}{\Gamma{(t+1)}}\approx \frac{T}{\Gamma{(1-T)}} \frac{1}{t^{1+T}}.
\end{align}

From~(\ref{eq:p_w_final}), the expected weight in the thermodynamic limit is
\begin{align}
\nonumber \bar{t}_{\textnormal{w}}  \xrightarrow{ N \to \infty } & \sum_{t=1}^{\tau} \frac{t}{w(\tau)} \frac{\Gamma{(\tau-t+T)}\Gamma{(t-T)}}{\Gamma{(\tau-t+1)}\Gamma{(t+1)}}\\
=& \frac{\Gamma{(1+T)} \Gamma{(\tau+1)}}{\Gamma{(\tau+T)}} \approx \Gamma{(1+T)} \tau^{1-T}.
\end{align}
The above relation decreases approximately exponentially with $T \in (0,1)$, and diverges at $\tau \to \infty$, 
\begin{align}
\bar{t}_{\textnormal{w}} \xrightarrow{\substack{N \to \infty \\ \tau \to \infty}} \infty.
\end{align}
We next turn our attention to the expected degree in the time-aggregated network.

%%%%%%%%%%%%%%%%%%%%%%%%%%%%%%%%%%%%%%%%%%%%%%%%%%%%%%%%%%%%%%%%%%%%
\subsection{Time-aggregated degree and finite size effects}
\label{sec:barkaggr}

The probability that two agents $i, j$ with latent variables $\kappa_i, \kappa_j$ do not interact, is obtained by setting $t=0$ in~(\ref{eq:uncon_w}),
\begin{align}
\label{eq:p_0_kappa_exact}
\nonumber r_\textnormal{w}(0; \kappa_i, \kappa_j) =& \frac{2 \mu \kappa_i \kappa_j }{N} \Bigg[\frac{T \Gamma{(\tau+T)}\Gamma{(-T)}}{\Gamma{(\tau)}}\\
+&(u_{ij}^\textnormal{min})^{-T} {}_2 F_{1}(-T, 1-\tau-T, 1-T, u_{ij}^\textnormal{min})\Bigg],
\end{align}
where $u_{ij}^\textnormal{min}$ in~(\ref{eq:u_min}). Removing the condition on $\kappa_i$ and $\kappa_j$ gives the probability that two agents do not interact
\begin{align}
\label{eq:r_0}
r_\textnormal{w}(0)=\int \int r_\textnormal{w}(0; \kappa_i, \kappa_j)  \rho(\kappa_i) \rho(\kappa_j) \mathrm{d}\kappa_i \mathrm{d} \kappa_j.
\end{align}

\begin{figure}
\includegraphics[width=2.75in]{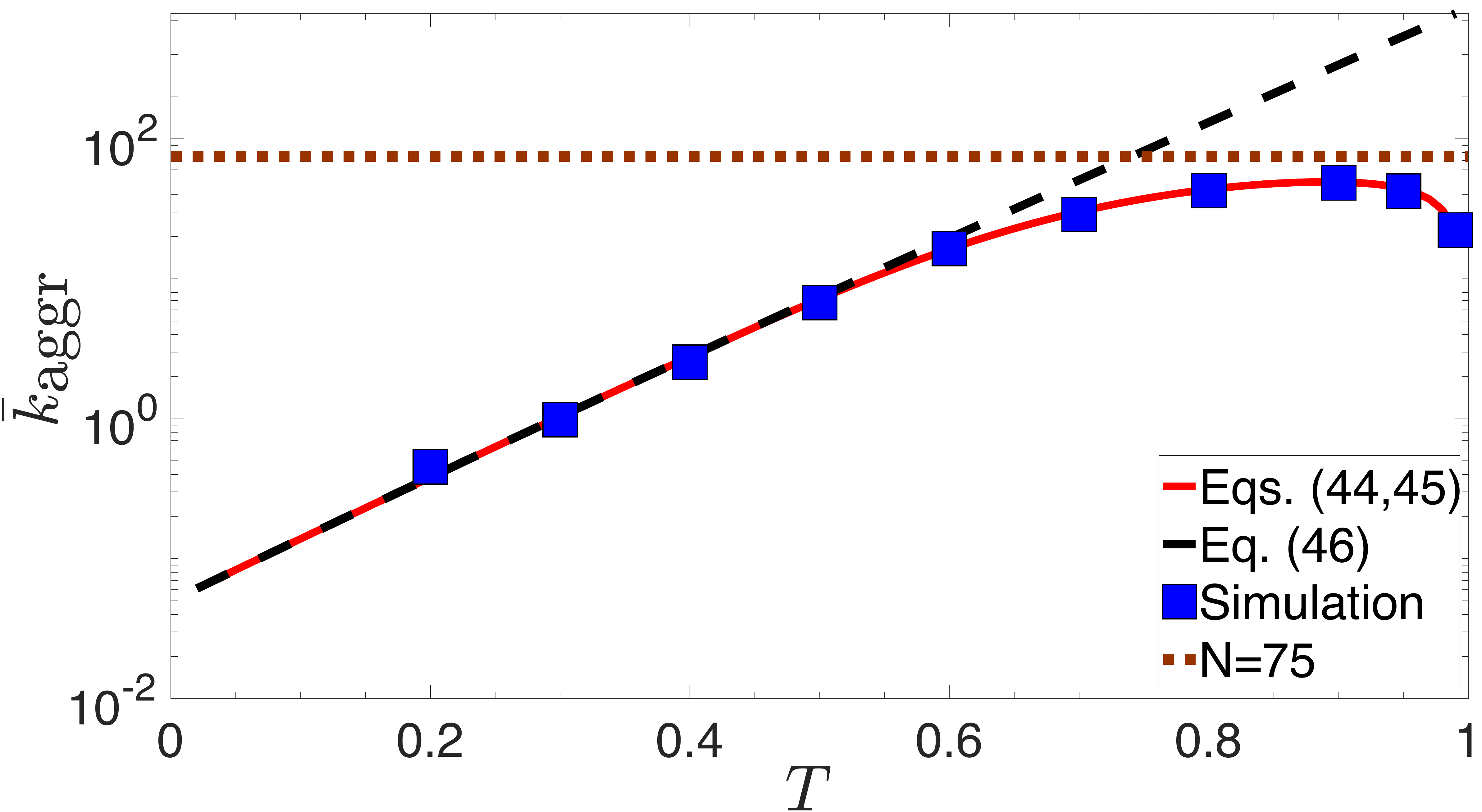}
\caption{Average time-aggregated degree as a function of the temperature $T$ in simulated networks~vs.~(\ref{eq:r_0}, \ref{eq:kaggr}) and~(\ref{eq:kaggr_limit}). The simulation parameters are $N=75,  \bar{k}=0.05$ and $\tau=17376$ (as in the hospital), while $\kappa_i=\bar{k}$, $\forall i$, i.e., the PDF of $\kappa$ is the Dirac delta function, $\rho(\kappa)=\delta(\kappa-\bar{k})$. 
\label{figbarkappa}}
\end{figure}

The expected time-aggregated degree is
\begin{align}
\label{eq:kaggr}
\bar{k}_{\textnormal{aggr}} = (N-1)\left[1-r_\textnormal{w}(0)\right].
\end{align}
At $N \to \infty$, $\bar{k}_{\textnormal{aggr}}$ is given by (\ref{eq:lim_4_uncon}). Substituting $\mu$ in (\ref{eq:lim_4_uncon}) with its expression in~(\ref{eq:mu}), gives
\begin{align}
\label{eq:kaggr_limit}
 \bar{k}_{\textnormal{aggr}}  \xrightarrow{ N \to \infty } \frac{\Gamma{(\tau+T)} \bar{\kappa}}{\Gamma{(1+T)}\Gamma{(\tau)}} \approx \frac{\tau^T\bar{\kappa}}{\Gamma{(1+T)}},
\end{align}
which increases exponentially with $T$ and linearly with $\bar{\kappa}$. Fig.~\ref{figbarkappa} juxtaposes simulation results against~(\ref{eq:r_0}, \ref{eq:kaggr}) and the limit in~(\ref{eq:kaggr_limit}). We see an excellent agreement between~(\ref{eq:r_0},  \ref{eq:kaggr}) and simulations, while~(\ref{eq:kaggr_limit}) is a good approximation only at sufficiently low temperatures. 

\begin{figure}
\includegraphics[width=2.8in]{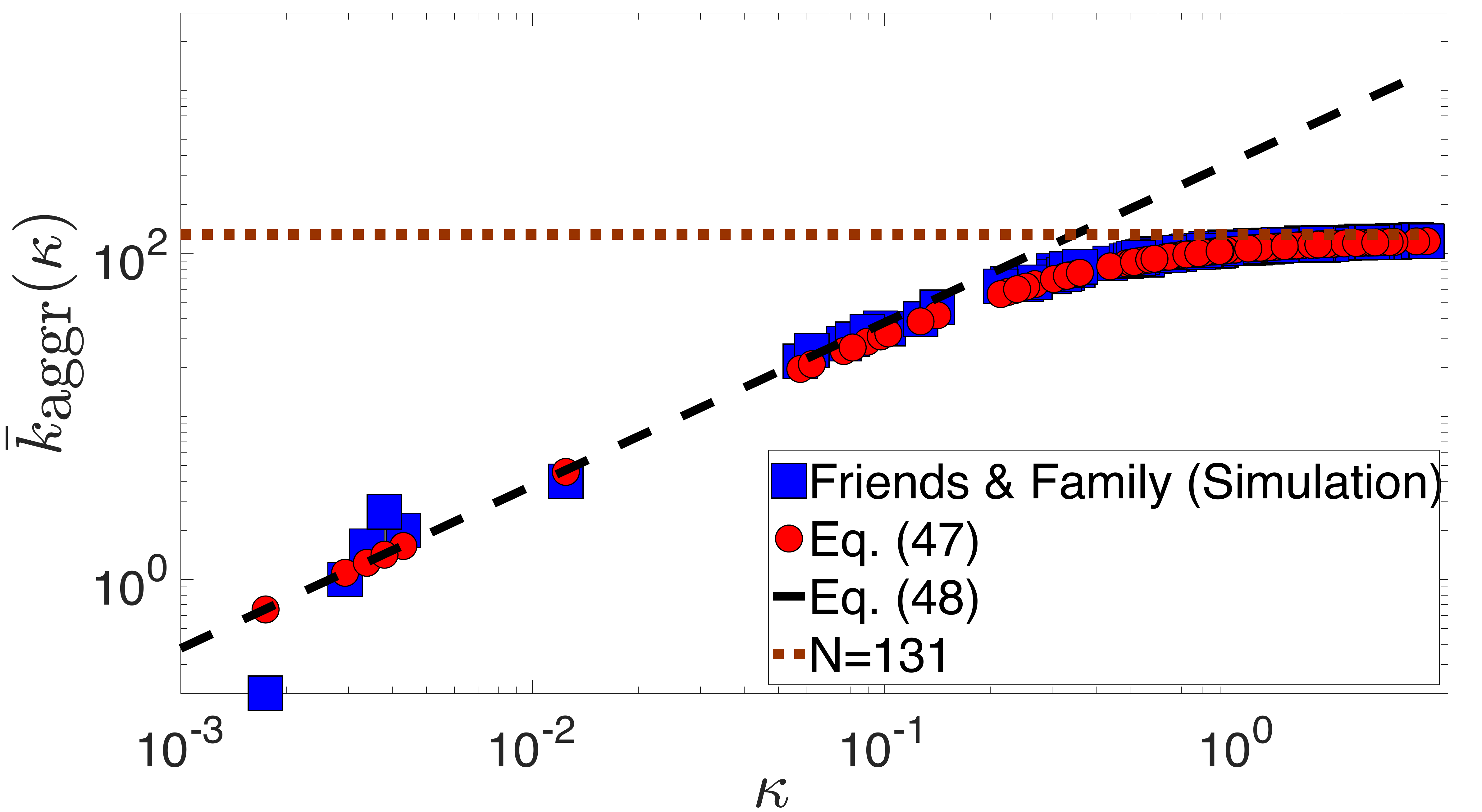}
\caption{Average time-aggregated degree as a function of the latent degree variable $\kappa$ in the simulated counterpart of the Friends \& Family (Sec.~\ref{sec:modeled_nets})~vs.~(\ref{eq:kaggr_kappa}) and~(\ref{eq:kaggr_kappa_limit}). The simulation results are averages over $5$ runs.
\label{fig:kaggr_k}}
\end{figure}

Similarly, the expected time-aggregated degree of a node with latent variable $\kappa_i$, is
\begin{align}
\label{eq:kaggr_kappa}
\bar{k}_{\textnormal{aggr}}(\kappa_i) &= (N-1)\left[1-\int r_\textnormal{w}(0; \kappa_i, \kappa_j) \rho(\kappa_j) \mathrm{d}\kappa_j \right]\\
\label{eq:kaggr_kappa_limit}
&\xrightarrow  { N \to \infty }  \frac{\Gamma{(\tau+T)} \kappa_i}{\Gamma{(1+T)} \Gamma{(\tau)}} \approx \frac{\tau^T \kappa_i}{\Gamma{(1+T)}}.
\end{align}
Fig.~\ref{fig:kaggr_k} juxtaposes simulation results against~(\ref{eq:kaggr_kappa}) and~(\ref{eq:kaggr_kappa_limit}). We again see an excellent agreement between the exact prediction~(\ref{eq:kaggr_kappa}) and simulations, while~(\ref{eq:kaggr_kappa_limit}) is a good approximation only for sufficiently small $\bar{k}_{\textnormal{aggr}}(\kappa)$. Therefore, one in general needs to use exact expressions [(\ref{eq:r_0}, \ref{eq:kaggr}), (\ref{eq:kaggr_kappa})] to accurately compute expected time-aggregated degrees. The thermodynamic limit approximations [(\ref{eq:kaggr_limit}), (\ref{eq:kaggr_kappa_limit})] are accurate only at sufficiently low temperatures.

\begin{figure}[!b]
\includegraphics[width=2.8in]{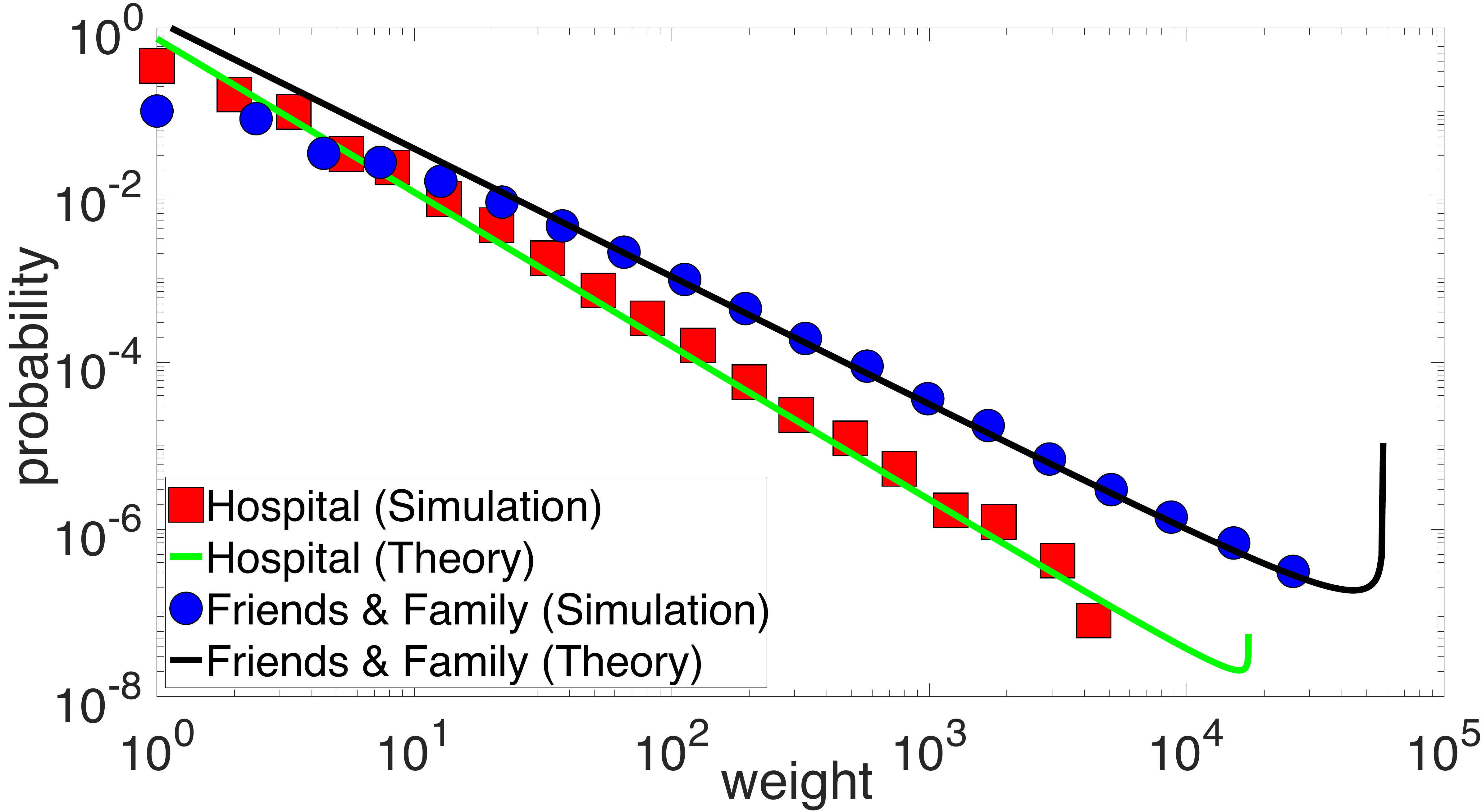}
\caption{Aggregated weight distribution in the simulated counterparts of the hospital and Friends \& Family (Sec.~\ref{sec:modeled_nets})~vs.~theoretical prediction given by~(\ref{eq:p_w_final}, \ref{eq:w_2}) with $\tau, T, \bar{k}_{\textnormal{aggr}}$ and $\bar{\kappa}=\bar{d}$ as in Table~\ref{tableSimulated}. The upward bendings at the tails of the distributions are due to the finite observation time $\tau$. Similar results hold for the rest of the counterparts.
\label{fig:validation3}}
\end{figure}

We also note that the normalization factor $w(\tau)$ of the weight distribution in~(\ref{eq:p_w_final}) can be rewritten as
\begin{align}
\label{eq:w_2}
w(\tau) = \frac{\Gamma{(1-T}) \Gamma{(T)} \bar{k}_{\textnormal{aggr}}}{\tau \bar{\kappa}},
\end{align}
where $\bar{k}_{\textnormal{aggr}}$ in~(\ref{eq:kaggr_limit}). Fig.~\ref{fig:validation3} juxtaposes~(\ref{eq:p_w_final}) against simulation results, where in view of Fig.~\ref{figbarkappa}, we use in~(\ref{eq:w_2}) the actual value of  $\bar{k}_{\textnormal{aggr}}$ in the simulations instead of its limit in~(\ref{eq:kaggr_limit}). We see again a very good agreement between theory and simulations.

%%%%%%%%%%%%%%%%%%%%%%%%%%%%%%%%%%%%%%%%%%%%%%%%%%%%%%%%%%%%%%%%%%%%
\subsection{Strength-degree correlations}
\label{sec:strength}

We now analyze the strength-degree correlations in the time-aggregated network and justify previous empirical observations reporting a super-linear dependence between an individual's expected strength and its time-aggregated degree~\cite{Starnini2013, StarniniDevices2017}.

\begin{figure}
\includegraphics[width=2.82in]{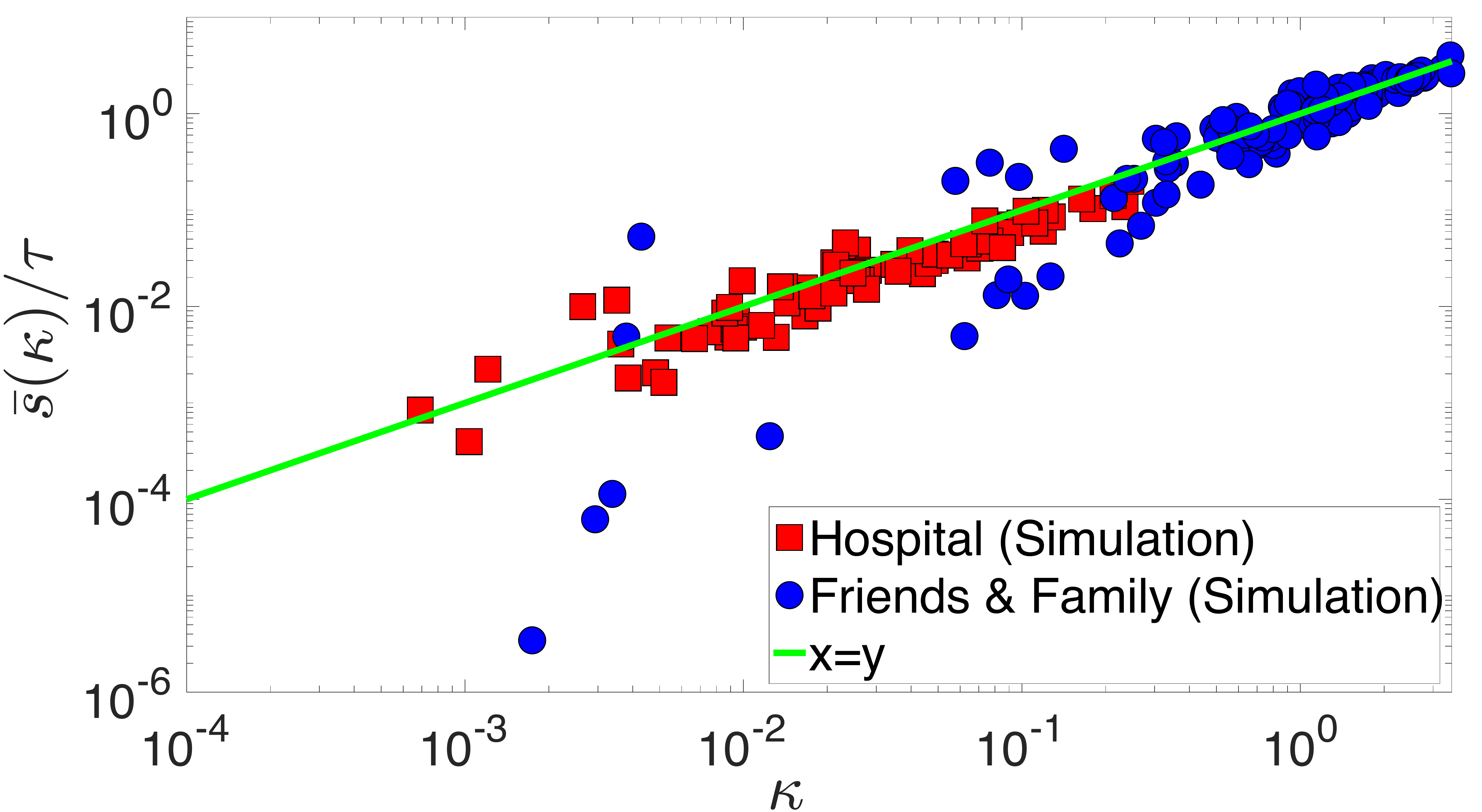}
\caption{Normalized average strength $\bar{s}(\kappa)/\tau$ as a function of the latent degree variable $\kappa$ in the simulated counterparts of the hospital and Friends \& Family (Sec.~\ref{sec:modeled_nets}). The results are averages over $20$ and $5$ runs, respectively. In the counterparts $\bar{k}=\bar{\kappa}$~($=\bar{d}$), canceling out in~(\ref{eq:s_k}).
\label{fig:s_k}}
\end{figure}

The expected weight between two nodes $i, j$ with latent variables $\kappa_i, \kappa_j$, is
\begin{align}
\overline{w}(\kappa_i, \kappa_j)=\sum_{t=1}^{\tau} t r_\textnormal{w}(t;\kappa_i, \kappa_j),
\end{align}
where $r_\textnormal{w}(t; \kappa_i, \kappa_j)$ in~(\ref{eq:uncon_w}). At $N \to \infty$, the second term inside the brackets in~(\ref{eq:uncon_w}) vanishes for $T \in (0, 1)$ and $t \geq 1$, yielding
\begin{align}
\label{eq:weight_limit}
\nonumber N \overline{w}(\kappa_i, \kappa_j) \xrightarrow { N \to \infty }& 2\mu \kappa_i \kappa_j T \tau \sum_{t=1}^{\tau} t \frac{\Gamma{(\tau-t+T)}\Gamma{(t-T)}}{\Gamma{(\tau-t+1)}\Gamma{(t+1)}}\\
=& \frac{\tau \bar{k} \kappa_i \kappa_j }{\bar{\kappa}^2}.
\end{align}
The expected strength of a node with latent variable $\kappa_i$, is
\begin{align}
\label{eq:s_k}
\bar{s}(\kappa_i)= N \int \overline{w}(\kappa_i, \kappa_j)\rho(\kappa_j) \mathrm{d}\kappa_j  \xrightarrow { N \to \infty } \frac{\tau \bar{k}  \kappa_i}{\bar{\kappa}}.
\end{align}
Fig.~\ref{fig:s_k} juxtaposes~(\ref{eq:s_k}) against simulation results. We see that~(\ref{eq:s_k}) can be a good approximation in finite networks. This is because the second term inside the brackets in~(\ref{eq:uncon_w}) vanishes even for finite networks as $t$ increases. The smaller the temperature the faster this term vanishes and the better the approximation in~(\ref{eq:s_k}) is for finite networks.

\begin{figure}
\includegraphics[width=1.68in, height=1.30in]{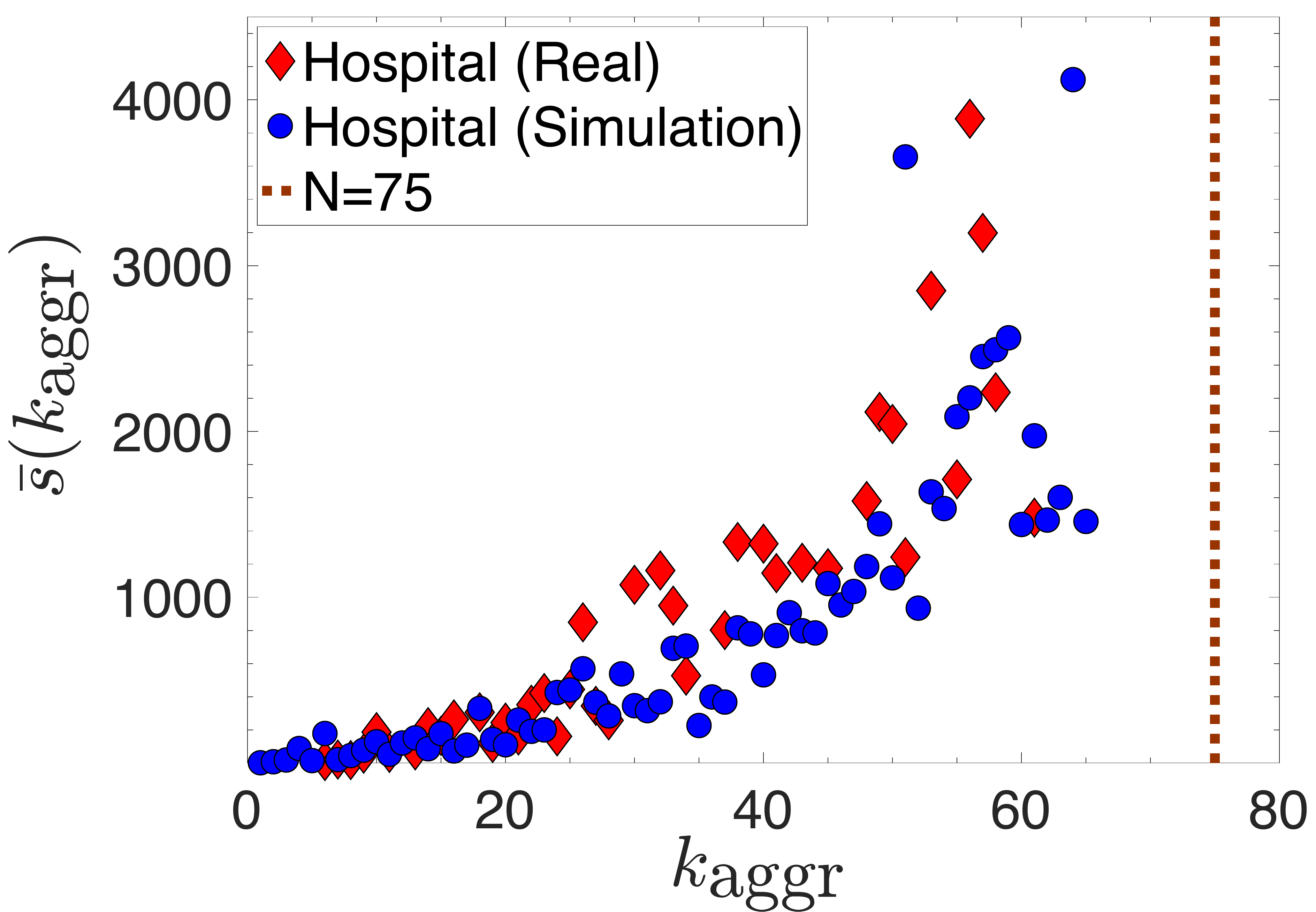}
\includegraphics[width=1.68in, height=1.38in]{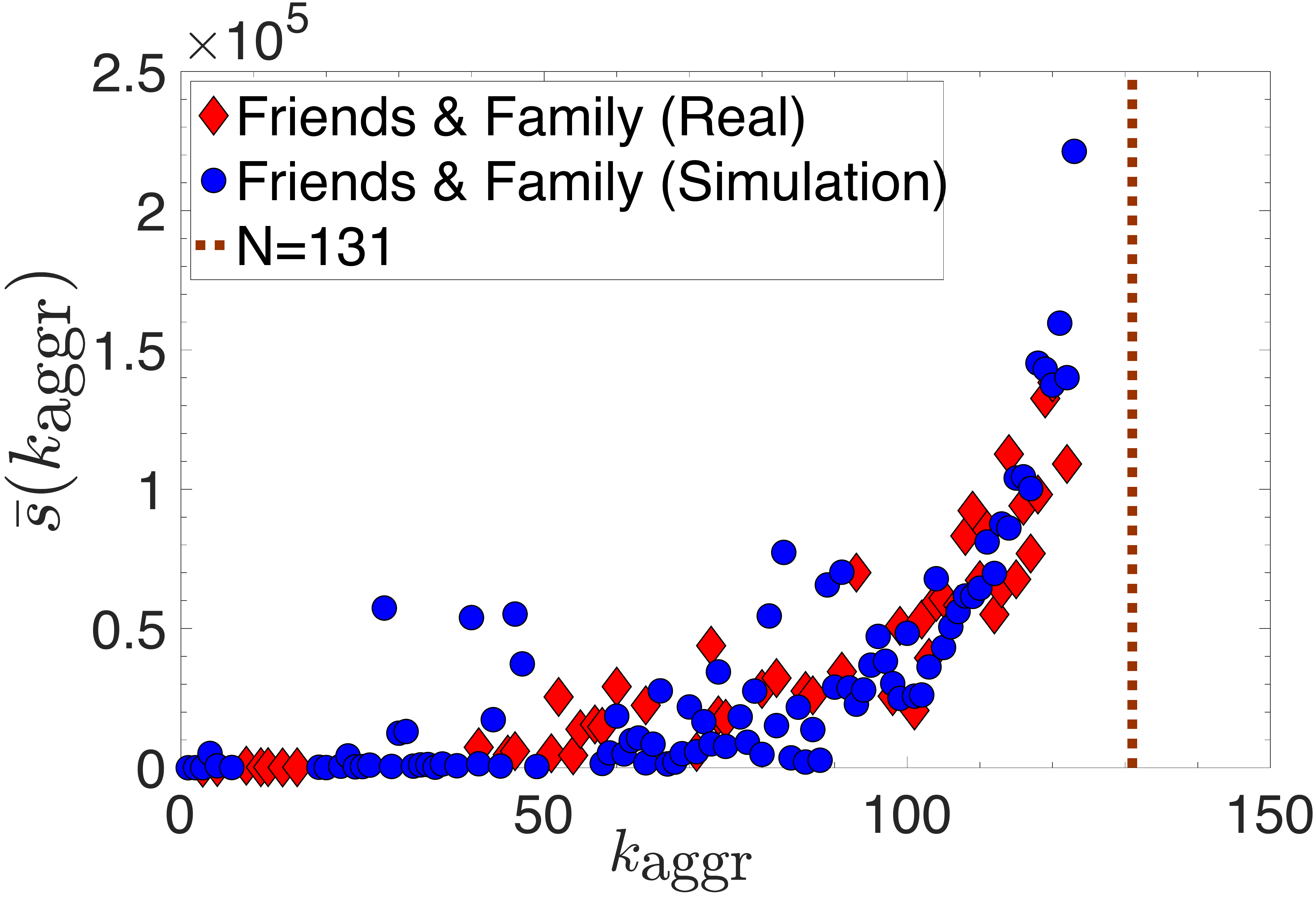}
\caption{Average strength as a function of the time-aggregated degree in real and simulated networks. Similar results hold for the rest of the real networks and their counterparts from Sec.~\ref{sec:modeled_nets}.
\label{fig:s_kaggr}}
\end{figure}

We also see from~(\ref{eq:kaggr_kappa_limit}, \ref{eq:s_k}) that in the thermodynamic limit the expected strength of a node grows linearly with its expected time-aggregated degree,
\begin{align}
\label{eq:s_vs_kaggr_limit}
\bar{s}(\kappa_i) \propto \bar{k}_{\textnormal{aggr}} (\kappa_i).
\end{align}
However, in the counterparts $\bar{k}_{\textnormal{aggr}} (\kappa_i)$ grows sub-linearly with $\kappa_i$ (Fig.~\ref{fig:kaggr_k}), while $\bar{s}(\kappa_i)$ grows approximately linearly (Fig.~\ref{fig:s_k}). Thus, in the considered systems we expect the strength of a node to grow super-linearly with its time-aggregated degree, as verified in Fig.~\ref{fig:s_kaggr} and empirically observed in prior studies~\cite{Starnini2013, StarniniDevices2017}.

%%%%%%%%%%%%%%%%%%%%%%%%%%%%%%%%%%%%%%%%%%%%%%%%%%%%%%%%%%%%%%%%%%%%
\section{Component dynamics and temperature}
\label{sec:components_vs_T}

\begin{figure}
\includegraphics[width=2.8in]{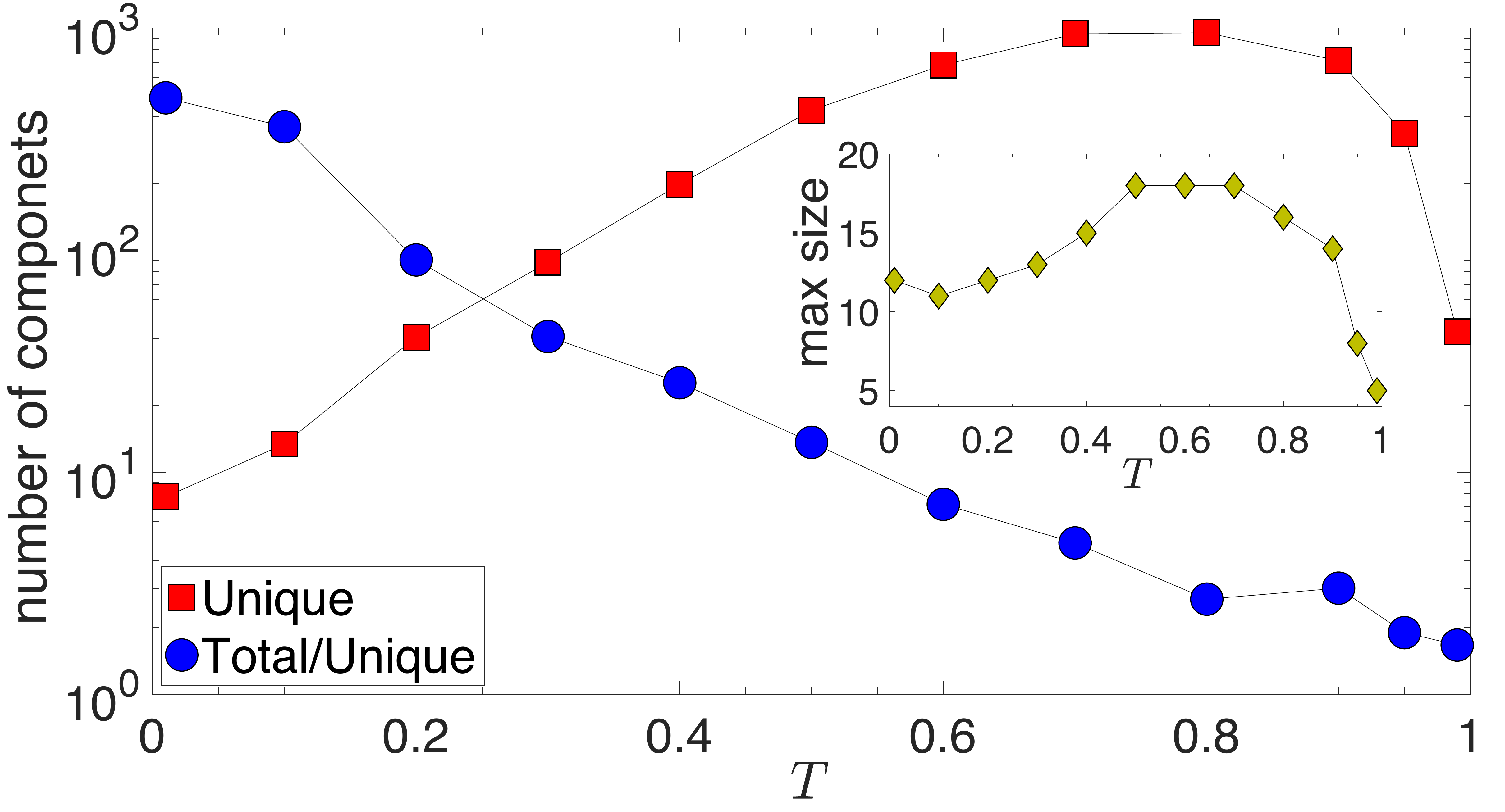}
\caption{Number and size of components formed~vs.~temperature $T$. The simulation parameters are the same as in the counterpart of the hospital (Sec.~\ref{sec:modeled_nets}) except that $T$ varies in $(0, 1)$. 
\label{fig:components_vs_T}}
\end{figure}

Finally, we elucidate the important role of the temperature $T$ in the formation of components. To this end, we consider the connected components formed in all time slots throughout the observation period $\tau$, which consist of at least three nodes. We consider both unique and recurrent components. A component in a slot is called \emph{unique} if it is seen for the first time, i.e., it is a component that does not consist of exactly the same nodes as a component seen in a previous slot. Otherwise, the component is recurrent. Fig.~\ref{fig:components_vs_T} shows that as $T$ increases, the number of unique components increases almost exponentially up to a point and then decreases.  This is because larger values of $T$ increase the connection probability [Eq.~(\ref{eq:p_s1})] at larger distances ($\chi_{ij} > 1$), while decreasing it at smaller distances ($\chi_{ij} < 1$). Since there are more pairs of nodes separated by larger distances, the number of unique components formed increases. However, at larger $T$ closer to one, the probability of connections is relatively small at smaller and larger distances, which causes this number to decrease. The inset in Fig.~\ref{fig:components_vs_T} shows the size of the largest component formed.

Further, Fig.~\ref{fig:components_vs_T} shows that the ratio of the total number of components formed to the number of unique components formed decreases with $T \in (0,1)$. This means that as $T$ increases fewer recurrent components are formed per unique component. This is expected since at larger $T$ unique components consist of pairs separated by larger distances, and the probability to form again the same such components is vanishing. 
\\

%%%%%%%%%%%%%%%%%%%%%%%%%%%%%%%%%%%%%%%%%%%%%%%%%%%%%%%%%%%%%%%%%%%%
\section{Conclusion}
\label{sec:conclusion}

Despite its simplicity the dynamic-$\mathbb{S}^{1}$ reproduces adequately many of the observed properties of real proximity networks.  At the same time the model is amenable to mathematical analysis. We have proved here the model's main properties (Sec.~\ref{sec:analysis}). Other properties were studied only via simulations (Sec.~\ref{sec:properties}) and it would be interesting in future work to prove those properties as well. We have seen that network temperature plays a central role in network dynamics, dictating the contact, inter-contact and weight distributions, the time-aggregated degrees, and the formation of unique and recurrent components.

The dynamic-$\mathbb{S}^{1}$ may not capture the properties of a real network \emph{exactly}. For instance, the aggregated contact, inter-contact and weight distributions may deviate from pure power laws, may follow power laws with exponential cutoffs, may have different exponents than exactly $2+T, 2-T, 1+T$, etc., cf.~Fig.~\ref{figFFProps}(a). Further, we have seen that the pairwise inter-contact distributions are on average more skewed in real networks than in the model. As future work, it would be also interesting to investigate what mechanisms need to be introduced into the model in order to be able to capture such variations.

We also note that memory in the dynamic-$\mathbb{S}^{1}$ is induced only via the nodes' latent variables ($\kappa, \theta$). Extensions to the model with \emph{link persistence}, where connections/disconnections can also be copied from the previous to the next snapshot~\cite{link_persistence_paper1, link_persistence_paper2}, would allow additional control over the rate of dynamics, i.e., on how fast the topology changes from snapshot to snapshot. Further, generalizations of the model that would allow the nodes' latent variables ($\kappa, \theta$) to change over time are desirable. However, for this purpose, one would first need to find the  equations that realistically describe the motion of nodes in their latent spaces. The dynamic-$\mathbb{S}^{1}$ or extensions of it may  apply to other types of time-varying networks, such as the ones considered in~\cite{Perra2012, Karsai2014}, and constitute the basis of maximum likelihood estimation methods that infer the node coordinates and their evolution in the latent spaces of real systems~\cite{KimSurvey}. Taken altogether, our results pave the way towards generative modeling of temporal networks that simultaneously satisfies simplicity, realism, and mathematical tractability.

%%%%%%%%%%%%%%%%%%%%%%%%%%%%%%%%%%%%%%%%%%%%%%%%%%%%%%%%%%%%%%%%%%%%
\begin{acknowledgments}
The authors acknowledge support by the EU H2020 NOTRE project (grant 692058).
\end{acknowledgments}
%%%%%%%%%%%%%%%%%%%%%%%%%%%%%%%%%%%%%%%%%%%%%%%%%%%%%%%%%%%%%%%%%%%%

\appendix
%%%%%%%%%%%%%%%%%%%%%%%%%%%%%%%%%%%%%%%%%%%%%%%%%%%%%%%%%%%%%%%%%%%%
\section{dynamic-$\mathbb{S}^{1}$ vs. configuration model}
\label{sec:scm}

The dynamic-$\mathbb{S}^{1}$ utilizes the $\mathbb{S}^{1}$ model at the cold regime where the temperature is $T < 1$ (Sec.~\ref{sec:S1}). The $\mathbb{S}^{1}$ can be also defined at the hot regime, $T > 1$~\cite{Krioukov2010}.

Like traditional complex networks~\cite{Krioukov2010}, proximity networks appear to belong to the cold regime. Indeed, as seen in Table~\ref{tableSimulated}, all counterparts have $T < 1$. Further, Fig.~\ref{fig:components_vs_T} shows that the number of recurrent components quickly decreases with $T \in (0,1)$, becoming small at $T \to 1$, while real networks have large numbers of recurrent components (cf.~Figs.~\ref{figAllProps}(h),~\ref{figFFProps}(h) and \cite{flores2018}).

Analyzing the dynamic-$\mathbb{S}^{1}$ at the hot regime is beyond the scope of this paper. However, we consider here a limiting case at this regime, where the $\mathbb{S}^{1}$ model degenerates to the configuration model, i.e., to the ensemble of graphs with given
expected degrees~\cite{ChungLu2002, Newman2004}. This case corresponds to letting $T \to \infty$, while completely ignoring the angular distances among the nodes, see~\cite{Krioukov2010} for details. The connection probability between two nodes $i, j$ becomes
\begin{align}
\label{eq:p_cm}
p_{\textnormal{cm}}(\kappa_i, \kappa_j)=\frac{1}{1+N \bar{\kappa}^2/(\bar{k} \kappa_i \kappa_j)}.
\end{align}
For sparse networks ($\bar{k} \ll N$) and distributions of $\kappa_i$ that are not too broad (conditions that hold in the considered networks,  Fig.~\ref{figKappas}), we can write $p_{\textnormal{cm}}(\kappa_i, \kappa_j) \approx \bar{k} \kappa_i \kappa_j/(N \bar{\kappa}^2)$. Using this approximation, it is easy to see that the expected degree of a node with latent variable $\kappa$ is given by~(\ref{eq:kappa}), while the average degree in the resulting network is $\bar{k}$.

\begin{figure*}
\includegraphics[width=18cm, height=15cm]{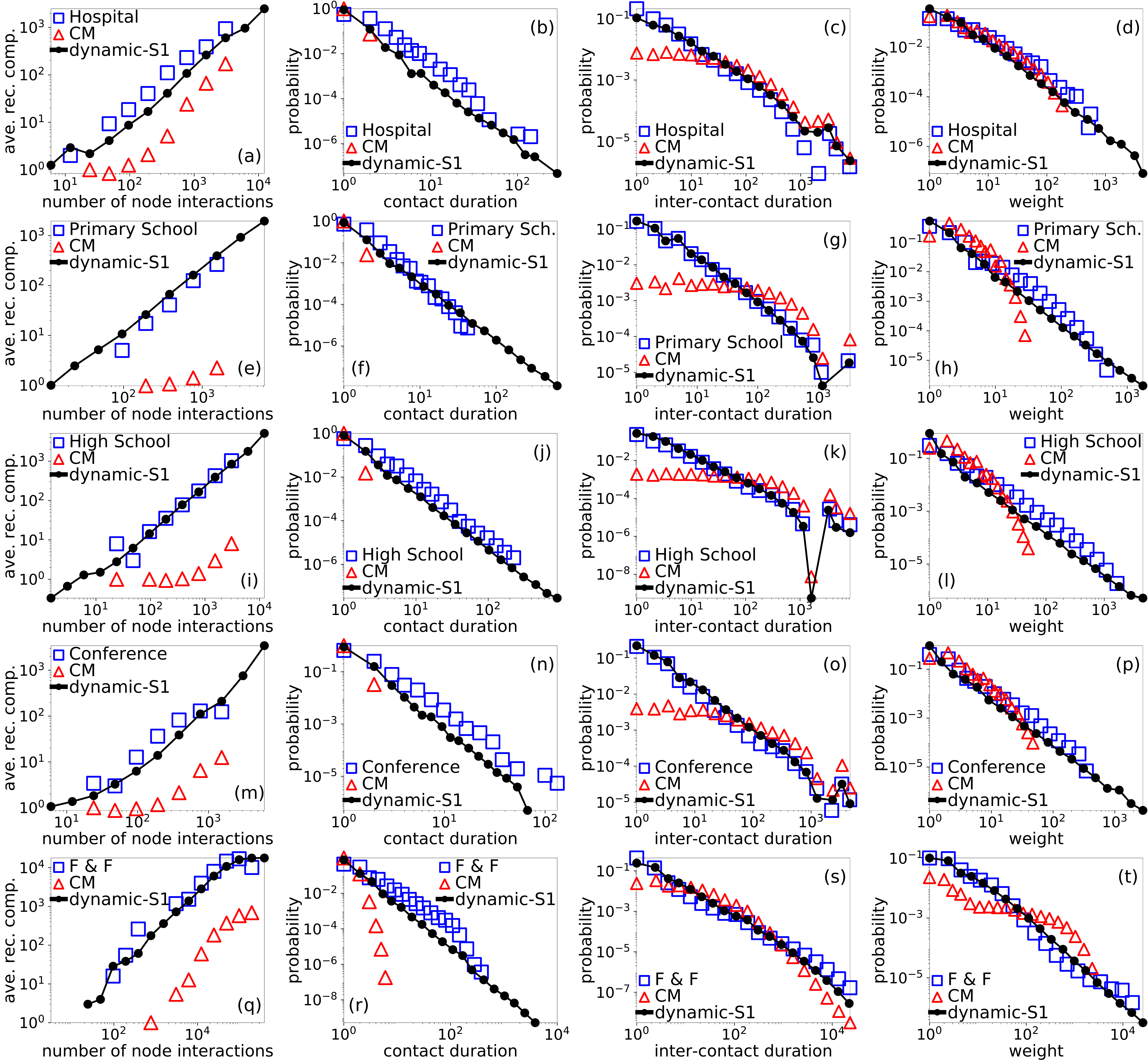}
\caption{Real face-to-face interaction networks vs. simulated networks with the configuration model (CM). 
\textbf{(a,e,i,m,q)}~Average number of recurrent components where an agent participates as a function of the total number of interactions of the agent.
\textbf{(b,f,j,n,r)}~Contact distribution. 
\textbf{(c,g,k,o,s)}~Inter-contact distribution. 
\textbf{(d,h,l,p,t)}~Weight distribution. 
For comparison the results with the dynamic-$\mathbb{S}^{1}$ considered in the main text are also shown. The results with the models are averages over $20$ simulation runs except from the Friends \& Family where the averages are over $5$ runs.
\label{figAllProps_cm}}
\end{figure*}

We now build synthetic counterparts for the real networks of Sec.~\ref{sec:real_nets} using the dynamic-$\mathbb{S}^{1}$ as described in Secs.~\ref{sec:dynamic_S1} and \ref{sec:modeled_nets}, except that we connect the nodes in each snapshot with the connection probability in~(\ref{eq:p_cm}) [instead of~(\ref{eq:p_s1})]. Since there is no temperature $T$ in (\ref{eq:p_cm}), we can no longer control the average time-aggregated degree, which is  significantly larger in the counterparts, $\bar{k}_\textnormal{aggr}=58, 214, 242, 76, 125$, for the hospital, primary school, high school, conference and Friends \& Family, respectively (vs. the ones in Table~\ref{tableReal}). As expected, we see in Fig.~\ref{figAllProps_cm} that the configuration model cannot reproduce the abundance of recurrent components observed in the real networks. Further, it cannot capture their broad contact, inter-contact and weight distributions (Fig.~\ref{figAllProps_cm}).

%%%%%%%%%%%%%%%%%%%%%%%%%%%%%%%%%%%%%%%%%%%%%%%%%%%%%%%%%%%%%%%%%%%%


\end{document}